\documentclass[aps,eqsecnum,preprint,floats,epsf,epsfig,nofootinbib]{revtex4}
\usepackage{graphicx}
\usepackage{epsfig}

\begin{document}
\def\be{\begin{eqnarray}}
\def\en{\end{eqnarray}}
\def\non{\nonumber}
\def\la{\langle}
\def\ra{\rangle}
\def\pp{{\prime\prime}}
\def\nc{N_c^{\rm eff}}
\def\vp{\varepsilon}
\def\hep{\hat{\varepsilon}}
\def\a{{\cal A}}
\def\B{{\cal B}}
\def\c{{\cal C}}
\def\d{{\cal D}}
\def\e{{\cal E}}
\def\P{{\cal P}}
\def\t{{\cal T}}
\def\up{\uparrow}
\def\dw{\downarrow}
\def\vma{{_{V-A}}}
\def\vpa{{_{V+A}}}
\def\smp{{_{S-P}}}
\def\spp{{_{S+P}}}
\def\J{{J/\psi}}
\def\3bar{{\bf \bar 3}}
\def\6bar{{\bf \bar 6}}
\def\10bar{{\bf \ov{10}}}
\def\ov{\overline}
\def\Lqcd{{\Lambda_{\rm QCD}}}
\def\pr{{Phys. Rev.}~}
\def\prl{{ Phys. Rev. Lett.}~}
\def\pl{{ Phys. Lett.}~}
\def\np{{ Nucl. Phys.}~}
\def\zp{{ Z. Phys.}~}
\def\lsim{ {\ \lower-1.2pt\vbox{\hbox{\rlap{$<$}\lower5pt\vbox{\hbox{$\sim$}
}}}\ } }
\def\gsim{ {\ \lower-1.2pt\vbox{\hbox{\rlap{$>$}\lower5pt\vbox{\hbox{$\sim$}
}}}\ } }

\font\el=cmbx10 scaled \magstep2{\obeylines\hfill May, 2004}

\vskip 1.5 cm

\centerline{\large\bf Light-Front Approach for Heavy Pentaquark
Transitions}
\bigskip
\centerline{\bf Hai-Yang Cheng,$^1$ Chun-Khiang Chua$^1$ and
Chien-Wen Hwang$^2$}
\medskip
\centerline{$^1$ Institute of Physics, Academia Sinica}
\centerline{Taipei, Taiwan 115, Republic of China}
\medskip

\medskip
\centerline{$^2$ Department of Physics, National Kaohsiung Normal
University} \centerline{Kaohsiung, Taiwan 802, Republic of China}
\medskip

\bigskip
\bigskip
\centerline{\bf Abstract}
\bigskip
\small

Assuming the two diquark structure for the pentaquark state as
advocated in the Jaffe-Wilczek model, there exist exotic
parity-even anti-sextet and parity-odd triplet heavy pentaquark
baryons. The theoretical estimate of charmed and bottom pentaquark
masses is quite controversial and it is not clear whether the
ground-state heavy pentaquark lies above or below the strong-decay
threshold. We study the weak transitions of heavy pentaquark
states using the light-front quark model. In the heavy quark
limit, heavy-to-heavy pentaquark transition form factors can be
expressed in terms of three Isgur-Wise functions: two of them are
found to be normalized to unity at zero recoil, while the third
one is equal to 1/2 at the maximum momentum transfer, in
accordance with the prediction of the large-$N_c$ approach or the
quark model. Therefore, the light-front model calculations are
consistent with the requirement of heavy quark symmetry. Numerical
results for form factors and Isgur-Wise functions are presented.
Decay rates of the weak decays $\Theta_b^+\to\Theta_c^0
\pi^+(\rho^+)$, $\Theta_c^0\to\Theta^+ \pi^-(\rho^-)$,
$\Sigma'^+_{5b}\to \Sigma'^0_{5c}\pi^+(\rho^+)$ and
$\Sigma'^0_{5c}\to N^+_8\pi^-(\rho^-)$ with $\Theta_Q$,
$\Sigma'_{5Q}$ and $N_8$ being the heavy anti-sextet, heavy
triplet and light octet pentaquarks, respectively, are obtained.
For weakly decaying $\Theta_b^+$ and $\Theta_c^0$, the branching
ratios of $\Theta_b^+\to\Theta_c^0\pi^+$,
$\Theta_c^0\to\Theta^+\pi^-$ are estimated to be at the level of
$10^{-3}$ and a few percents, respectively.

\eject
\section{Introduction}

The discovery of an exotic $\Theta^+$ pentaquark by LEPS at
SPring-8 \cite{LEPS}, subsequently confirmed by many other groups
\cite{Diana,CLAS1,Saphir,ITEP,CLAS2,Hermes,SVD,COSY,ZEUS}, marked
a new era for testing our understanding of the hadron spectroscopy
and promoted a re-examination of the QCD implications for exotic
hadrons. The mass of the $\Theta^+$ is of order 1535 MeV and its
width is less than 10 MeV from direct observations and can be as
narrow as 1 MeV from the analysis of $K$-deuteron scattering data
\cite{width}. Another exotic pentaquark $\Xi^{--}_{3/2}$ with a
mass of $1862\pm 2$ MeV and a width smaller than 18 MeV was
observed by NA49 \cite{NA49}. In spite of the confirmation of the
$\Theta^+$ from several experiments, the pentaquark candidate
signals must be established beyond any doubt as all current
experimental signals are weak and the significance is only of 4--6
standard deviations. Indeed, there are several null results of
pentaquark search from e.g., BES \cite{BES} and HERA-B
\cite{HeraB}. For example, the reaction $pA\to \Xi^\pm\pi^\mp$ at
$\sqrt{s}=41.7$ GeV has been studied by HERA-B and no narrow
signal in the $\Xi^\pm\pi^\mp$ invariant mass distribution is
seen. The $\Theta^+$ pentaquark is also not observed by NA49, BES
and HERA-B.

In the conventional uncorrelated quark model, the $\Theta^+$ mass
is expected to be of order 1900 MeV for a $S$-wave ground state
with odd parity and 2200 MeV for a $P$-wave state with even
parity. The width is at least of order 100 MeV as the strong decay
$\Theta^+\to KN$ is OZI super-allowed. Therefore, within the naive
uncorrelated quark model one cannot understand why $\Theta^+$ is
anomalously light and why its width is so narrow. This hints a
possible correlation among various quarks; two or three quarks
could form a cluster. Several quark cluster models have been
proposed in the past. Jaffe and Wilczek \cite{JW} advocated a two
diquark picture in which the $\Theta^+$ is a bound state of an
$\bar s$ quark with two $(ud)$ diquarks. The diquark is a highly
correlated spin-zero object and is in a flavor anti-triplet and
color anti-triplet state. The parity of $\Theta^+$ is flipped from
the negative, as expected in the naive quark model, to the
positive owing to the diquark correlation. The even parity of the
$\Theta^+$ is in agreement with the prediction of the chiral
soliton model \cite{DPP}. In the model of Karliner and Lipkin
\cite{KL}, $\Theta^+$ is composed of the diquark $(ud)$ and the
triquark $(ud\bar s)$. Under the {\it assumption} that the diquark
and triquark are in a relative $P$-wave, the resultant $\Theta^+$
also has a positive parity. Note that two of previous lattice
calculations imply a negative parity for the
$\Theta^+$~\cite{Csikor,Sasaki}. However, based on the
Jaffe-Wilczek picture to construct the interpolating operators, a
recent quenched lattice QCD calculation with exact chiral symmetry
yields a positive parity for the pentaquark states $\Theta^+$ and
$\Xi^{--}_{3/2}$ \cite{Chiu}.

A major distinction between the Jaffe-Wilczek model and the chiral
soliton model is that the pentaquark baryons including $\Theta^+$
in the latter model are in the pure $\10bar_f$ representation of
the flavor SU(3), while the former has not only anti-decuplet but
also octet even-parity pentaquark baryons. The nearly ideal mixing
of $\10bar_f$ and ${\bf 8}_f$ in the Jaffe-Wilczek model leads to
the prediction that the pentaquark $\Xi_{3/2}$ is lighter than
$\Sigma_\10bar$, opposite to the chiral soliton model where
$\Xi_{3/2}$ is the heaviest one among the anti-decuplet pentaquark
baryons. Hence, a measurement of the $\Sigma_\10bar$ mass can be
used to discriminate between the chiral soliton model and any
quark model with ideal mixed pentaquark states.

Although the Jaffe-Wilczek model is simple, powerful and leads to
many concrete predictions, it has its own difficulties: (i)
Assuming that the octet pentaquark state with nucleon quantum
numbers can be identified with the Roper resonance $N(1440)$, the
predicted $\Xi_{3/2}$ mass of 1750 MeV is smaller than the NA49
measurement by an amount of 110 MeV. (ii) The predicted mass of
the anti-decuplet $N_\10bar$ with nucleon quantum numbers is
around 1700 MeV. It is tempted to identify this with
$N(1710)P_{11}$ as suggested by Jaffe and Wilczek. However, as
pointed out in \cite{Cohen}, the identification of the two nucleon
pentaquark states as the Roper resonance and the $N(1710)$ is
inconsistent with the phenomenology of the widths given the
assumptions underlying the model and the narrow $\Theta^+$ width:
The broad Roper resonance will imply a large partial width of
$N(1710)$. This means that the model requires the existence of at
least one presently unknown penqaquark state with nucleon quantum
numbers and a mass in the neighborhood of 1700 MeV. (iii) The
model also predicts a negative-parity nonet of pentaquark states
which are supposed to be lighter than the even-party octet
pentaquark baryons owing to the lack of the $P$-wave excitation
energy. However, except for the SU(3)-singlet $\Lambda(1405)$
which may be a pentaquark state as indicted by recent lattice
calculations \cite{Lambda1405}, the observed spectroscopy of the
${1\over 2}^-$ states $N(1535),\Sigma(1620),\Lambda(1670)$ is
known to be well described as the orbital excited states of
three-quark baryons. The question is then why the new ${1\over
2}^-$ octet pentaquark states have not been found in the energy
regime which has been well explored. (iv) Nussinov  has pointed
out that a strict point-like diquark picture of the $\Theta^+$
conflicts with QCD inequalities \cite{Nussinov}. This suggests
that the picture of a point-like diquark scalar may miss
significant hyperfine interactions between the $\bar s$ and each
of the four quarks in the pentaquark. This will be elaborated more
in Sec. II.B.

Given the existence of the $\Theta^+$ pentaquark, it is natural to
consider its heavy flavor analogs $\Theta_c^0$ and $\Theta_b^+$ by
replacing the $\bar s$ quark in $\Theta^+$ by the heavy antiquark
$\bar c$ and $\bar b$, respectively. Whether the masses of the
heavy pentaquark states are above or below the strong-decay
threshold has been quite controversial. Very recently, a narrow
resonance in $D^{*-}p$ and $D^{*+}\bar p$ invariant mass
distributions was reported by the H1 Collaboration \cite{H1}. It
has a mass of $3099\pm3\pm5$ MeV and a Gaussian width of $12\pm3$
MeV and can be identified with the spin 1/2 or 3/2 charmed
pentaquark baryon. Although this state is about 300 MeV higher
than the $DN$ threshold, it is possible that the observed H1
pentaquark is a chiral partner of the yet undiscovered ground
state $\Theta_c^0$ with opposite parity and a mass of order 2700
MeV as implied by several model estimates \cite{Nowak} (see Sec.
II.B for more details). The latter pentaquark can be discovered
only through its weak decay.

In the Jaffe-Wilczek model, there exist parity-even antisextet and
parity-odd triplet heavy pentaquarks containing a single heavy
antiquark $\bar c$ or $\bar b$ and they are all truly exotic. The
heavy pentaquark baryons in the $\3bar_f$ representation are
lighter than the $\6bar_f$ ones due to the lack of orbital
excitation and therefore may be stable against strong decays
\cite{Stewart}. Consequently, it becomes important to study the
weak and electromagnetic decays of heavy pentaquarks.

It is well known that the study of nonleptonic weak decays of
baryons is much more complicated than the meson case for several
reasons. First, the baryon is made of three quarks and hence there
exist many more quark diagrams responsible for the weak decays of
baryons. Second, the factorizable approximation that the hadronic
matrix element is factorized into the product of two matrix
elements of single currents and that the nonfactorizable term such
as the $W$-exchange contribution is negligible relative to the
factorizable one works empirically  reasonably well for describing
the hadronic weak decays of heavy mesons. However, this
approximation is {\it a priori} not directly applicable to heavy
baryon decays. For example, the $W$-exchange diagram is no longer
subject to helicity and color suppression. Indeed, it is known
that $W$ exchange plays an essential role in describing the data
of charmed baryon decays.

At first sight, it is natural to expect a much more complicated
dynamics for the nonleptonic weak decays of pentquark baryons.
Nevertheless, the underlying mechanism is greatly simplified if
the penatquark can be approximately described by the diquark
picture of Jaffe and Wilczek. In this scenario it is equivalent to
working on the effective three-body problem. Moreover, in the
heavy quark limit, pentaquark transitions can be described in
terms of several universal Isgur-Wise functions. The point is that
in the infinite quark mass limit, the heavy quark spin $S_Q$
decouples from the other degrees of freedom of the hadron, so that
$S_Q$ and the total angular momentum $j$ of the light quarks are
separately good quantum numbers. Whether the total spin of the
light degrees of freedom comes from the quark spin plus orbital
angular momentum or just from the quark spin is irrelevant. As a
result, the previous studies of heavy-to-heavy baryon transitions
using heavy quark effective theory in nineties can be easily
generalized to the heavy-to-heavy pentaquark transitions. The
heavy quark limit results are model independent and hence must be
respected by any explicit model calculations.

In the present work we shall study the pentaquark transitions
using the relativistic light-front quark model and assuming the
two diquark structure for the pentaquarks as described by the
Jaffe-Wilczek model. The light-front model allows us to study the
transition form factors and their momentum dependence.
Furthermore, large relativistic effects which may manifest near
the maximum large recoil, i.e. $q^2=0$, are properly taken into
account in the light-front framework. In the literature the weak
decays of heavy pentaquark states have been studied in
\cite{Leibovich,Stewart,He}.

The layout of the present paper is organized as follows. In Sec.
II we discuss the mass spectrum of heavy pentaquark baryons in the
diquark picture. In Sec. III we present a detailed study of the
pentaquark weak transitions within the light-front quark model and
derive the analytic expressions for form factors. The heavy quark
limit behavior of the form factors is studied in Sec. IV and the
universal Isgur-Wise functions are obtained. Numerical results for
form factors, Isgur-Wise functions and examples of weak decays of
heavy pentaquark baryons are worked out in Sec. V. Conclusion is
given in Sec. VI.

\section{Diquark Model for Heavy Pentaquarks}
\subsection{Jaffe-Wilczek model}

In the pentaquark model of Jaffe and Wilczek \cite{JW}, the heavy
flavor pentaquark consists of a heavy antiquark $\bar Q$ and two
diquark pairs $[q_1q_2]$ and $[q_3q_4]$. The diquark $[q_1q_2]$ is
a highly correlated spin-zero object and is in a flavor
anti-triplet $\3bar_f$ and color anti-triplet $\3bar_c$
representation. The two diquarks form a flavor anti-sextet
$\6bar_f$ or triplet ${\bf 3}_f$ configuration. The diquark pair
must be in color ${\bf 3}_c$ state in order to form a
color-singlet pentaquark. Bose statistics of the scalar diquarks
requires that the diquark pairs in the flavor symmetric $\6bar_f$
(antisymmetric ${\bf 3}_f$) state be in an orbital $P$ ($S$) wave
or a spatially antisymmetric (symmetric) state. Since the heavy
quark is an SU(3) flavor-singlet, it is clear that heavy
pentaquarks form an SU(3)-flavor antisextet with even parity and
triplet with odd parity. The members of $\6bar_f$ pentaquarks
containing an anti-charmed quark are $\Theta_c^0,
\Sigma_{5c}^{0,-}$ and $\Xi_{5c}^{0,-,--}$, while members of ${\bf
3}_f$ are $\Sigma'^{0,-}_{5c}$ and $\Xi'^-_{5c}$ (we follow
\cite{Huang} for the notation).\footnote{The notation is opposite
to the case of charmed baryons where the unprimed states
$\Sigma_c$ and $\Xi_c$ are for ${\bf 6}_f$ and the primed states
$\Sigma'_c$ and $\Xi'_c$ are for $\3bar_f$. In \cite{Stewart}, the
charm and bottom pentaquarks are denoted by $T_a$ and $R_a$,
respectively. More precisely, $T_s^{0,-},T_{ss}^-$
($R_s^{+,0},R_{ss}^0$) correspond to our
$\Sigma'^{0,-}_{5c},\Xi'^-_{5c}$
($\Sigma'^{+,0}_{5b},\Xi'^0_{5b}$).} The corresponding flavor wave
functions are summarized in Table I. Note that the resulting $J^P$
of $\6bar_f$ pentaquarks is either $1/2^+$ or $3/2^+$, while $J^P$
is equal to ${1/2}^-$ for ${\bf 3}_f$ heavy penaquarks. Naively,
the ${\bf 3}_f$ pentaquark $\P_Q(\bf 3)$ is expected to be lighter
$\P_Q(\6bar)$ owing to the absence of the $P$-wave orbital
excitation. However, this may be compensated by the Pauli blocking
repulsion as we shall discuss below in more detail.

In the Karliner-Lipkin model \cite{KL,KLheavy}, the pentaquark is
composed of a triquark $q_1q_2\bar Q$ and a diquark $[q_3q_4]$.
The two quarks $q_1q_2$ in the triquark are in ${\bf 6}_c$ and
$\3bar_f$ representations, while the spin zero diquark $[q_3q_4]$
is in $\3bar_c$ and $\3bar_f$ configuration. Consequently, we have
$\6bar_f$ and ${\bf 3}_f$ heavy pentaquarks. However, unlike the
Jaffe-Wilczek model, Karliner and Lipkin {\it assumed} a $P$-wave
orbital angular momentum between the two clusters. As a
consequence, the resulting ${\bf 3}_f$ heavy penaquark has
$J^P=1/2^+$ and is degenerate with the $\6bar_f$ one. Therefore, a
study of the parity of triplet heavy pentaquarks will enable us to
discriminate between Jaffe-Wilczek and Karliner-Lipkin models.

\begin{table}[h]
\caption{Flavor wave functions of heavy pentaquarks in the
Jaffe-Wilczek model, where $\bar Q=\bar c$ or $\bar b$ and
$[q_1q_2][q_3q_4]_\pm=\sqrt{1\over 2}([q_1q_2][q_3q_4]\pm
[q_3q_4][q_1q_2])$.} \label{}
\begin{ruledtabular}
\begin{tabular}{ c c | c c }
 State & ~~~~~~~Flavor wave function~~~~~~~~~~~~~ & State & Flavor wave function \\
 \hline
 $\Theta_c^0,\Theta_b^+$ & $[ud]^2\bar Q$ & & \\
 $\Sigma_{5c}^0,\Sigma_{5b}^+$ & $[ud][us]_+\bar Q$ &
 $\Sigma'^0_{5c},\Sigma'^+_{5b}$ & $[ud][us]_-\bar Q$ \\
 $\Sigma_{5c}^-,\Sigma_{5b}^0$ & $[ud][ds]_+\bar Q$ &
 $\Sigma'^-_{5c},\Sigma'^0_{5b}$ & $[ud][ds]_-\bar Q$ \\
 $\Xi^0_{5c},\Xi^+_{5b}$ & $[us]^2\bar Q$ & & \\
 $\Xi^-_{5c},\Xi^0_{5b}$ & $[us][ds]_+\bar Q$ &
 $\Xi'^-_{5c},\Xi'^0_{5b}$ & $[us][ds]_-\bar Q$ \\
 $\Xi_{5c}^{--},\Xi_{5b}^-$ & $[ds]^2\bar Q$ & & \\
\end{tabular}
\end{ruledtabular}
\end{table}

For light pentaquarks in the Jaffe-Wilczek model, the diquark pair
$[q_1q_2][q_3q_4]$ is combined with a light antiquark to form a
pentaquark baryon. When the diquark pair is in the
flavor-symmetric $\6bar_f$ representation, the flavor content of
the resulting even-parity pentaquark states is ${\bf 8}_f\oplus
\ov{\bf 10}_f$. When $q_1\neq q_2$ and $q_3\neq q_4$, the three
diquark pairs $[ud][us],[ds][su]$ and $[su][ud]$ can be
antisymmetrized in flavor and hence they are allowed to have
symmetric, positive parity, spatial wavefunctions. Therefore,
there is a non-exotic nonet of pentaquark baryons with negative
parity and flavor content ${\bf 3}_f\otimes\3bar_f={\bf 1}_f\oplus
{\bf 8}_f$. It is naively expected that the ${1\over 2}^-$ nonet
is lighter than the ${1\over 2}^+$ octet pentaquarks owing to the
lack of orbital excitation. However, since the two diquarks in the
former are in a relative $S$ wave, they have substantial overlap
at short distances. Consequently, the identical particles in
different diquarks (e.g. the $u$ quark in $[us][ud]$) will
experience a repulsive interaction due to Pauli blocking
\cite{JW}. Jaffe and Wilczek conjectured that this effect is
strong enough to elevate the masses of the resultant negative
parity pentaquarks to the range of 1.5--2.0 GeV.\footnote{By
neglecting the Pauli blocking effect, the masses of the ${1\over
2}^-$ pentaquark nonet are estimated in \cite{Zhang} to lie in the
range of 1360 to 1540 MeV.}
These baryons are expected to be very board. The negative-parity
baryon resonances that have been observed are the nonet states
$N(1535),\Sigma(1620),\Lambda(1670),\Lambda(1405)$ and the octet
states $N(1650),\Sigma(1750),\Lambda(1800)$ \cite{PDG}. As stated
in the Introduction, $N(1535),\Sigma(1620),\Lambda(1670)$ are
known to be well described as the orbital excited states of
three-quark baryons, while recent lattice calculations indicate
that the SU(3)-singlet $\Lambda(1405)$ could be a pentaquark state
\cite{Lambda1405}. Hence, it is not clear if the second octet
states $N(1650),\Sigma(1750),\Lambda(1800)$ can be identified as
${1\over 2}^-$ pentaquarks. At any rate, it seems plausible to
argue that the Pauli blocking effect is at least comparable to the
$P$-wave excitation energy.

\subsection{Masses of heavy pentaquark states}
We first consider the masses of $\Theta_b^+$ and $\Theta_c^0$.
Arguing that the $[ud]$ diquark in the $\Lambda_c$ and $\Lambda$
experiences nearly the same environment as in $\Theta_c^0$ and
$\Theta^+$, respectively, Jaffe and Wilczek estimated that
\cite{JW}
 \be \label{eq:ThetamassJW}
 m(\Theta_c^0) &=& m(\Theta)+m(\Lambda_c)-m(\Lambda)=2710\,{\rm MeV}, \non \\
 m(\Theta_b^+) &=& m(\Theta)+m(\Lambda_b)-m(\Lambda)=6050\pm10\,{\rm
 MeV}.
 \en
Assuming the color-spin interaction as the dominant mechanism for
hyperfine splitting, Cheung obtained \cite{Cheung}
 \be \label{eq:ThetamassCheung}
 m(\Theta_c^0)=(2938-2997)\,{\rm MeV}, \qquad
 m(\Theta_b^+)=(6370-6422)\,{\rm MeV}
 \en
in the Jaffe-Wilczek model, while Karliner and Lipkin found
\cite{KLheavy}
  \be \label{eq:ThetamassKL}
 m(\Theta_c^0)=2985\pm 50\,{\rm MeV}, \qquad
 m(\Theta_b^+)=6389\pm 50\,{\rm MeV}
 \en
in their diquark-triquark model. It appears that the latter two
calculations yield similar results. Since the threshold for strong
decays into $DN$ and $BN$ is 2805 and 6217 MeV, respectively, it
is clear that the strong decays $\Theta_c^0\to D^-p$ and
$\Theta_b^+\to B^0p$ will be kinematically allowed if the
$\Theta^0_c$ and $\Theta_b^+$ masses are those given by
(\ref{eq:ThetamassCheung}) or (\ref{eq:ThetamassKL}). However, it
should be stressed that the color-spin hyperfine interaction
formula employed by Cheung or by Karliner and Lipkin is applicable
in principle only to $S$-wave hadronic systems \cite{Cheung}.
Moreover, it is not clear if the color-spin interaction is the
dominant hyperfine splitting effect.

It is worth mentioning the estimate of the $\Theta_c^0$ mass in
other calculations. In the chiral soliton model, the heavy
pentaquark has been described as a bound state of the SU(2) chiral
soliton and a heavy meson in terms of the bound state approach of
\cite{Callan}. The predicted $\Theta_c^0$ mass lies below the $DN$
threshold \cite{Oh} and is consistent with the estimate by Jaffe
and Wilczek. More recently, the average $\Theta_c$ mass, defined
by $\ov m(\Theta_c)=[m(\Theta_c)+2m(\Theta_c^*)]/3$, is estimated
based on the generalized SU(4) collective-coordinate quantization
in Skyrme model \cite{MaThetac}. It yields 2704 MeV and is again
very close to the estimate of (\ref{eq:ThetamassJW}). An early
calculation in a Goldstone boson exchange model also obtained a
${\Theta_c}$ mass below the strong decay threshold~\cite{Stancu}.
In contrast, a lattice calculation by Sasaki~\cite{Sasaki} leads
to $m(\Theta_c^0)=3445$ MeV, which is 640 MeV above the $DN$
threshold. Another quenched lattice QCD calculation with exact
chiral symmetry predicts a mass of $2977\pm109$~MeV for the
$\Theta_c^0$ \cite{privateChiu}.

At first sight, it appears that the aforementioned controversy
about the charmed pentaquark mass will be settled down by the
recent H1 observation of a narrow resonance in $D^{*-}p$ and
$D^{*+}\bar p$ invariant mass distributions which can be
identified with the spin 1/2 $\Theta_c^0$ and/or spin 3/2
$\Theta_c^{*0}$~\cite{H1}. However, as pointed out in
\cite{Nowak}, it is possible that the H1 state $\Theta_c^0(3099)$
is a chiral partner of another yet undiscovered ground state
pentaquark $\Theta_c^0(2700)$ with opposite parity. In this case,
$\Theta_c^0(2700)$ can be discovered only by studying its weak
decays. Therefore, it is important to measure the parity of the H1
state $\Theta_c^0(3099)$. If it has an odd parity, this may imply
the existence of a parity-even charmed pantaquark baryon with a
mass below the $DN$ threshold. In this case, one can follow the
argument of Jaffe and Wilczek to estimate the masses of other
heavy pentaquark baryons to be
 \be
 m(\Sigma_{5c}^0) &= &m(N_{\overline{10}}^+)+{1\over 2}\Big[
 m(\Lambda_c)-m(\Lambda)+m(\Xi_c)-m(\Xi)\Big]=2860\,{\rm MeV}, \non \\
 m(\Sigma_{5b}^+) &=& m(N_{\overline{10}}^+)+{1\over 2}\Big[
 m(\Lambda_b)-m(\Lambda)+m(\Xi_b)-m(\Xi)\Big]=6199\,{\rm MeV}, \non \\
 m(\Xi_{5c}^0) &=& m(\Xi_{3/2})+m(\Xi_c)-m(\Xi)=3014\,{\rm MeV}, \non \\
 m(\Xi_{5b}^0) &=& m(\Xi_{3/2})+m(\Xi_b)-m(\Xi)=6351\,{\rm MeV},
 \en
where use of $m(N_{\overline{10}})=1700$ MeV \cite{JW},
$m(\Xi_{3/2})=1862$ MeV \cite{NA49} and $m(\Xi_b)=5804$ MeV
\cite{Jenkins} has been made. Note that our numerical results are
slightly different than that
 in \cite{Huang}.

If the H1 state turns out to have a positive parity and is a truly
ground-state charmed pentaquark, it will have an important
implication for the Jaffe-Wilczek model, namely, the diquark
should not be treated as an idealized point-like scalar and there
are significant hyperfine attractive interactions between the
light antiquark and other four quarks of the pentaquark state
\cite{Nussinov}. The argument goes as follows. The observed
$\Theta_c^0$ mass can be accounted for in the uncorrelated quark
model in which the constituent quark masses of the charmed quark
and the light $u$ or $d$ quark are of order 1700 and 350 MeV,
respectively. However, the narrow width of $12\pm3$ MeV cannot be
explained by this naive model as the pentaquark  is in the
$S$-wave state. The $\Theta_c^0$ width will be suppressed if its
parity is even: The spatial separation between the quarks due to
the centrifugal barrier arising from the orbital angular momentum
will help reduce the decay width. (Of course, small widths are not
necessary the consequence of the centrifugal barrier.) Since the
$P$-wave excitation energy is of order 300 MeV (see below), this
means that the effective mass of the diquark $[ud]$ in the
Jaffe-Wilczek model is about 550 MeV. Presumably, the hyperfine
interaction of each quark of the diquark pair with the $\bar c$ is
small. However, in order to account for the $\Theta^+$ mass, the
effective mass of the diquark should be only of order 400 MeV.
This implies that for the light pentaquark $\Theta^+$, each quark
of the diquark pair has significant hyperfine attractive
interactions with the $\bar s$. In the constituent quark picture,
this is equivalent to stating that the correlation for the $ud$
pair is stronger in $\Theta^+$ than in $\Theta_c^0$. It is worth
remarking that the hyperfine splitting arising from the
interaction of the diquark pair with the antiquark is found to be
proportional to $-m_u/m_s$ and $-m_u/m_c$ for $\Theta^+$ and
$\Theta_c^0$, respectively, in \cite{Cheung}.

We next turn to the triplet penatquark $\P_Q({\bf 3})$ and the
spin 3/2 antisextet pentaquark $\P_Q^*(\6bar)$. In heavy quark
effective theory (HQET), the mass of a baryon containing a single
heavy quark has the $1/m_Q$ expansion \cite{Neubert94}
 \be
 m(H_Q)=m_Q+\bar\Lambda_{H_Q}-{\lambda_1\over 2m_Q}-d_H{\lambda_2\over
 2m_Q}+{\cal O}\left({1\over m_Q^2}\right),
  \en
where
  \be
 \lambda_1 &=& \la H_Q(v)|\bar Q_v(iD)^2Q_v|H_Q(v)\ra, \non \\
 d_H\lambda_2 &=& \la H_Q(v)|\bar Q_vg
 \sigma_{\mu\nu}G^{\mu\nu}Q_v|H_Q(v)\ra.
 \en
The three nonperturbative HQET parameters $\bar\Lambda_{H_Q},~
\lambda_1$, $\lambda_2$ in above equations are independent of the
heavy quark mass and $\bar\Lambda_{H_Q}$, the mass of the light
degrees of freedom of the hadron, in general varies for different
heavy hadrons. Since $\sigma\cdot G\sim \vec{S}_Q\cdot \vec{B}$
and since the chromomagnetic field is produced by the light cloud
inside the hadron, it is clear that $\sigma\cdot G$ is
proportional to $\vec{S}_Q\cdot\vec{S}_\ell$. Hence, the Clebsch
factor $d_H$ has the expression
 \be
d_H &=& -\la H_Q|4\vec{S}_Q\cdot\vec{S}_\ell|H_Q\ra   \non \\
&=& -2[S_{\rm tot}(S_{\rm tot}+1)-S_Q(S_Q+1)-S_\ell(S_\ell+1)],
 \en
where $\vec{S}_Q~(\vec{S}_\ell)$ is the spin operator of the heavy
quark (light cloud). Since $S_\ell=1$ (0) for the $\P_Q(\6bar)$
($\P_Q({\bf 3}))$ states, it follows that $d_H=0$ for triplet
pentaquark states, $d_H=4$ for spin-${1\over 2}$ antisextet
pentaquarks and $d_H=-2$ for the spin-${3\over 2}$ antisextet
pentaquark baryons. For a study of $\bar\Lambda_{H_Q}$ in terms of
$1/N_c$ expansion, see \cite{Jenkins}.

We first discuss the mass of the spin 3/2 pentaquark
$\P_Q^*(\6bar)$. It has the expression
 \be \label{eq:m3/2}
 m(\P^*_Q(\6bar))=m(\P_Q(\6bar))+{6\lambda_2^{\rm penta}\over
 2m_Q}.
 \en
Although we do not know the magnitude of $\lambda_2^{\rm penta}$
for the pentaquark system, the normal heavy baryons can provide a
useful guideline. The parameter $\lambda_2^{\rm baryon}$ can be
extracted from the mass splitting between ${3\over 2}^+$ and
${1\over 2}^+$ sextet heavy baryons, for example \cite{PDG}
 \be
 m[\Sigma_c(3/2^+)]-m[\Sigma_c(1/2^+)]\approx m[\Xi_c(3/2^+)]-m[\Xi_c(1/2^+)]
 \approx 65\,{\rm MeV}.
 \en
The result is \cite{Cheng97a}
 \be \label{eq:lambda2}
\lambda_2^{\rm baryon}=\cases{ 0.055\,{\rm GeV}^2 & for~charmed
baryons; \cr 0.041\,{\rm GeV}^2 & for~$\Sigma_b$; \cr  0.040\,{\rm
GeV}^2 & for~$\Xi'_b$; \cr  0.039\,{\rm GeV}^2 & for~$\Omega_b$.
\cr  }
 \en
We see that the heavy quark symmetry requirement that $\lambda_2$
be independent of the heavy quark mass is fairly respected. From
Eqs. (\ref{eq:m3/2}) and (\ref{eq:lambda2}) we find
 \be
 m(\P^*_c(\6bar))=m(\P_c(\6bar))+65\,{\rm MeV}, \qquad
 m(\P^*_b(\6bar))=m(\P_b(\6bar))+20\,{\rm MeV},
 \en
when $\lambda_2^{\rm penta}$ is set to $\lambda_2^{\rm baryon}$.

As for the triplet  pentaquark state $\P_Q({\bf 3})$, it is
naively expected to be lighter than $\P_Q(\6bar)$ by the amount of
the orbital excitation energy, which can be roughly estimated from
the mass difference between various ${1\over 2}^-$ and ${1\over
2}^+$ baryons to be \cite{PDG}:
 \be
 m[\Lambda(1/2^-)]-m[\Lambda(1/2^+)] &=& 290\,{\rm MeV},
 \non \\ m[\Lambda_c(1/2^-)]-m[\Lambda_c(1/2^+)] &=& 308.9\pm 0.6\,{\rm MeV},
 \\
 m[\Xi^+_c(1/2^-)]-m[\Xi^0_c(1/2^+)] &=& 318.2\pm3.2\,{\rm MeV},
 \non \\
 m[\Xi^0_c(1/2^-)]-m[\Xi^+_c(1/2^+)] &=& 324.0\pm3.3\,{\rm MeV}.
 \non
 \en
However, one should also take into account the Pauli blocking
effect and the intrinsic mass difference between ${\bf 3}_f$ and
$\6bar_f$. Denoting $\delta P$ and $\delta B$ as the $P$-wave
excitation energy and Pauli blocking repulsion, respectively, the
masses of the triplet heavy pentaquarks have the expressions:
 \be
 m(\P'_Q)=m(\P_Q)+\delta M_Q-\delta P_Q+\delta
 B_Q
 \en
where $\delta P_Q\approx 310$ MeV is suggested by the charmed
baryon data, and the quantity
 \be
 \delta M_Q=\bar\Lambda_{\bf 3}^{\rm penta}-\bar\Lambda_\6bar^{\rm penta}
 +{4\lambda_2^{\rm penta}\over  2m_Q}
 \en
is basically the mass difference between spin 1/2 triplet and
antisextet pentaquarks with the same parity. It is shown in
\cite{Jenkins} that an $1/N_c$ expansion of $\bar\Lambda$ yields
$\bar\Lambda=c_0N_c+c_2(S_\ell^2/N_c)$ and hence
$\bar\Lambda_\6bar >\bar\Lambda_{\bf 3}$. For normal charmed
baryons, $\bar\Lambda_\6bar^{\rm baryon}-\bar\Lambda_{\bf 3}^{\rm
baryon}=144$ MeV  inferred from the measured mass difference of
$\Xi'_c$ and $\Xi_c$: $\delta M_c^{\rm
baryon}=m(\Xi_c)-m(\Xi'_c)=-107$ MeV \cite{PDG}. This then leads
to $\delta M_b^{\rm baryon}\approx -131$ MeV. Presumably, $\delta
M^{\rm penta}_Q$ is not far from $\delta M^{\rm baryon}_Q$ for
heavy pentaquark baryons.

By neglecting the Pauli blocking effect and setting $\delta
M_Q^{\rm penta}=0$, it was argued in \cite{Stewart} that
$\P_Q({\bf 3})$ is stable against strong decays. We see that
whether the triplet pentaquark is stable against strong decays
depends on the Pauli blocking repulsion, $\delta M_Q$ and the mass
of $\P_Q(\6bar)$. Nevertheless, it is most likely that the triplet
pentaquark is lighter than the antisextet one unless the Pauli
blocking effect is unusually large.

\subsection{Decays of heavy pentaquark states}
The heavy pentaquarks can be studied via their strong,
electromagnetic and weak decays as discussed below.

(i) strong decays:
 \be
 (a) && \P_Q^{(*)}(\6bar)\to \B({\bf 8})+M_Q,  \non \\
 (b) && \P_Q^{(*)}(\6bar)\to \P_Q({\bf 3})+M, \\
 (c) && \P_Q({\bf 3})\to \B({\bf 8})+M_Q, \non
 \en
where $\B$ stands for the normal baryon made of three quarks.
Examples in this category are $\Xi_{5c}^0\to \ov
D^0\Xi^0,D_s^-\Sigma^+$, $\Sigma^{*0}_{5c}\to \Sigma'^-_{5c}\pi^+$
and $\Sigma'^0_{5c}\to\ov D^0\Sigma^0,D^-\Sigma^+,D_s^-p$. Of
course, whether the above-mentioned strong decays are
kinematically allowed or not depends on the heavy pentaquark
masses.

(ii) electromagnetic decays:
 \be
 (d) && \P_Q^*(\6bar)\to \P_Q(\6bar)+\gamma, \non \\
 (e) && \P_Q(\6bar)\to \P_Q({\bf 3})+\gamma.
 \en
The decays $\Sigma_{5c}^{*}\to\Sigma_{5c}\gamma$,
$\Sigma_{5c}\to\Sigma'_{5c}\gamma$ and $\Xi_{5c}\to
\Xi'_{5c}\gamma$ are examples. These radiative decays are
kinematically allowed.

(iii) weak decays:
 \be
 (f) && \P_b(\6bar)\to \P_c(\6bar)+M, \non \\
 (g) && \P_b(\6bar)\to \P_c({\bf 3})+M, \non \\
 (h) && \P_Q(\6bar)\to \B({\bf 8})+M,  \\
 (i) && \P_c(\6bar)\to \P_s(\10bar)+M, \non \\
 (j) && \P_b({\bf 3})\to \P_c({\bf 3})+M, \non \\
 (k) && \P_Q({\bf 3})\to \B({\bf 8})+M. \non
 \en
Examples are $\Theta_b^+\to\Theta_c^0\pi^+$,
$\Theta_c^0\to\Theta^+\pi^-$, $\Sigma'^+_{5b}\to
\Sigma'^0_{5c}\pi^+$, $\Sigma'^0_{5c}\to N_\10bar^+\pi^-$,
$\Xi'^-_{5c}\to\Lambda\pi^-$. In addition to nonleptonic decays,
there are also semileptonic weak decays.

Heavy pentaquark states can be produced in weak decays of heavy
hadrons, in $e^+e^-$ annihilation and at hadronic colliders
\cite{Rosner,Armstrong,Browder}. If the $\P_Q$ lies above the
strong-decay threshold, it will appear as a narrow $M_Q\B$
resonance. If pentaquarks lie below the threshold, then $\P_Q({\bf
3}),\Theta_Q,\Xi_{5c}^{0,--},\Xi_{5b}^{+,-}$ will have only weak
decays, while $\Sigma_Q,\Xi_{5c}^-,\Xi_{5b}^0$ will be dominated
by the electromagnetic interactions. In the above listed decays,
the strong decay (b) can be studied using the soft pion theorem.
The strong coupling constant involved in this process can be
related through PCAC to the matrix element of the axial-vector
current between the initial and final heavy pentaquark states with
the same heavy flavor. The weak decay modes (f), (g), (i) and (j)
receive factorizable contributions which can be expressed as the
product of the meson decay constant and the matrix element between
the initial and final heavy pentaquark states with different heavy
flavors. The process (g) will be vanished if the meson is emitted
from the $b\to c$ transition. The diquarks of the pentaquark
behave as spectators in this category of weak decays. The decays
of the pentaquark to ordinary octet baryons (h) and (k), e.g.
$\Sigma'^0_{5c}\to\Lambda K^0$ and $\Xi'^-_{5c}\to\Xi^-K^0$,
proceed via nonfactorizable $W$ exchange and hence the diquark
correlation is broken. Other hadronic decay channels of $\P_Q({\bf
3})$ can be found in \cite{Stewart}. It is interesting to notice
that the E791 Collaboration has searched for the triplet
pentaquark $\Sigma_{5c}'^0$ via the decays $\Sigma'^0_{5c}\to
\phi\pi p$ and $K^{*0}K^-p$ with null result \cite{E791}.

\section{Formalism of a light-front model for pentaquarks}

\begin{figure}[t!]
\centerline{
            {\epsfxsize4.5 in \epsffile{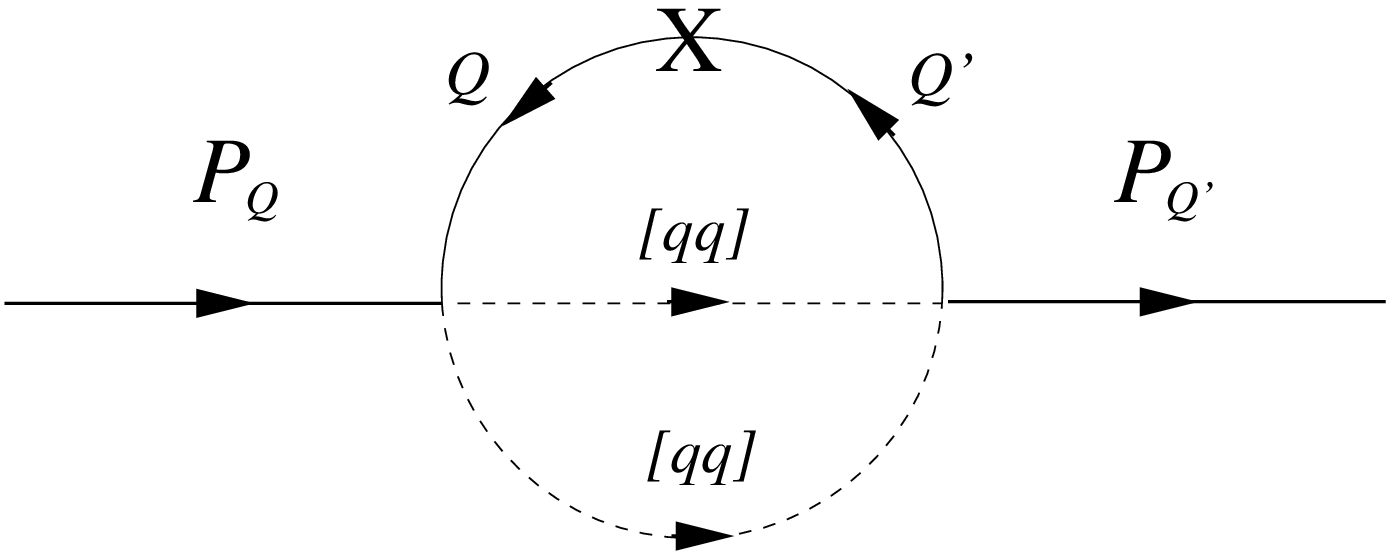}}}
\caption{Feynman diagram for a typical $\P_Q\to\P_{Q'}$
transition, where the spin-zero diquarks $([qq]=[ud],[us],[ds])$
are denoted by dashed lines and the corresponding $V-A$ current
vertex by $X$.} \label{fig:penta} 
\end{figure}

As stressed in Sec. II, the theoretical estimate of charmed and
bottom pentaquark masses and the issue of whether the ground-state
heavy pentaquark lies above or below the strong-decay threshold
are quite controversial. Even if the H1 state $\Theta_c^0(3099)$
is confirmed, the existence of a ground-state charmed pentaquark
with opposite parity and smaller mass is not ruled out. In this
section we shall focus on the hadronic weak decays of heavy
pentaquarks and study the pentaquark weak transitions within the
light-front approach and the Jaffe-Wilczek model. As mentioned
above, the hadronic two-body weak decays of a heavy pentaquark
receive factorizable contributions if the final-state baryon is a
pentaquark. This is not the case for a pentaquark decaying to an
ordinary baryon. Therefore, we will concentrate on $\P_Q\to
\P_{Q'}$ weak transitions where the diquark structure is
maintained. The corresponding figure is shown in
Fig.~\ref{fig:penta}.

\subsection{Vertex functions in the light-front approach}

We adopt the Jaffe-Wilczek picture~\cite{JW} for the heavy
pentaquark $\P_Q$ which has the quark flavor content ${\bar Q}[q_1
q_2][q_3 q_4]$.  For the purpose of the calculational convenience,
we shall treat the antiquark $\bar Q$ as a particle $Q^c$ instead
of an antiparticle, i.e. we use the charge conjugated field for
the heavy flavor. The reason of this seemingly odd choice will
become clear in later calculations. Note that although we consider
$\P_Q$, the formalism developed in this section can be easily
generalized to light pentaquark states, especially, for
$\Theta_s^+\sim \bar s[ud][ud]$.

In the light-front approach, the pentaquark bound state with the
total momentum $P$, spin $J=1/2$ and the orbital angular momentum
of the diquark pair $L=0,1$ can be written as (see, for example
\cite{Cheng97,CCH})
\begin{eqnarray}
        |\P_Q(P,L,S_z)\rangle
                =\int &&\{d^3p_1\}\{d^3p_2\}\{d^3p_3\} ~\frac{2(2\pi)^3}{\sqrt {P^+}} \delta^3(
                \tilde P -\tilde p_1-\tilde p_2-\tilde p_3)~\nonumber\\
        &&\times \sum_{\lambda_1,m,\alpha-\epsilon,b-e}
                \Psi^{S_z}_L(x_1,x_2,x_3,k_{1\bot},k_{2\bot},k_{3\bot},\lambda_1)~
                C^\alpha_{\beta\gamma,\delta\epsilon} F^{bc,de}_L
        \non\\
        &&\times ~
             \Big|Q^c_\alpha(p_1,\lambda_1) [q_b^\beta q_c^\gamma](p_2) [q_d^\delta q_e^\epsilon](p_3)\Big\rangle,
 \label{lfmbs}
\end{eqnarray}
where $\alpha,\cdots,\epsilon$ and $b,\cdots,e$ are color and
flavor indices, respectively, $\lambda$ denotes helicity, $p_1$,
$p_2$ and $p_3$ are the on-mass-shell light-front momenta,
\begin{equation}
        \tilde p=(p^+, p_\bot)~, \quad p_\bot = (p^1, p^2)~,
                \quad p^- = {m^2+p_\bot^2\over p^+},
\end{equation}
and
\begin{eqnarray}
        &&\{d^3p\} \equiv {dp^+d^2p_\bot\over 2(2\pi)^3},
        \quad \delta^3(\tilde p)=\delta(p^+)\delta^2(p_\bot),
        \nonumber \\
        &&\Big|Q^c(p_1,\lambda_1) [q_b q_c](p_2) [q_d q_e](p_3)\Big\rangle
        = d^\dagger_{\lambda_1}(p_1) \frac{a^\dagger(p_2) a^\dagger(p_3)}{\sqrt2}|0\rangle,\\
        &&[a(p'),a^\dagger(p)] =2(2\pi)^3~\delta^3(\tilde p'-\tilde
        p),\,
        \{d_{\lambda'}(p'),d_{\lambda}^\dagger(p)\} =
        2(2\pi)^3~\delta^3(\tilde p'-\tilde p)~\delta_{\lambda'\lambda}.
                \nonumber
\end{eqnarray}
The coefficient $C^\alpha_{\beta\gamma,\delta\epsilon}$ is a
normalized color factor and $F^{bc,de}_L$ is a normalized flavor
coefficient, obeying the relation
 \be
 &&C^{\alpha'}_{\beta'\gamma',\delta'\epsilon'} F^{b'c',d'e'}_{L}
 C^\alpha_{\beta\gamma,\delta\epsilon} F^{bc,de}_L
         \Big \la Q^{'c}_{\alpha'}(p'_1,\lambda'_1)
                 [q_{b'}^{\beta'} q_{c'}^{\gamma'}](p'_2)
                 [q_{d'}^{\delta'} q_{e'}^{\epsilon'}](p'_3)
             \Big|Q^c_\alpha(p_1,\lambda_1) [q_a^\beta q_b^\gamma](p_2) [q_c^\delta q_d^\epsilon](p_3)
             \Big\rangle
 \non\\
&&=2^3(2\pi)^9~\delta^3(\tilde p'_1-\tilde p_1)\frac{1}{2}[
 \delta^3(\tilde p'_2-\tilde p_2)\delta^3(\tilde p'_3-\tilde p_3)
 +(-)^L\delta^3(\tilde p'_2-\tilde p_3)\delta^3(\tilde p'_3-\tilde
 p_2)]\delta_{\lambda'_1\lambda_1}.
 \label{eq:norm}
 \en
Note that $C^\alpha_{\beta\gamma,\delta\epsilon}F^{bc,de}_L$ is
(anti-)symmetric under
$(\beta\gamma,bc)\leftrightarrow(\delta\epsilon,de)$ for $L=1~(0)$
as discussed in Sec.~II.A. As we shall see below, the factor of
$(-)^L$ will be compensated by the corresponding wave function
under the $p_2\leftrightarrow p_3$ interchange.

In terms of the light-front relative momentum variables $(x_i,
k_{i\bot})$ for $i=1,2,3$ defined by
\begin{eqnarray}
        && p^+_i=x_i P^{+}, \quad \sum_{i=1}^3 x_i=1, \nonumber \\
        && p_{i\bot}=x_i P_\bot+k_{i\bot}, \quad \sum_{i=1}^3 k_{i\bot}=0,
\end{eqnarray}
the momentum-space wave-function $\Psi^{S_z}_L$ can be expressed
as
\begin{equation}
        \Psi^{S_z}_L(x_i,k_{i\bot},\lambda_1,m)
                = \langle \lambda_1|{\cal R}_M^\dagger(x_1,k_{1\bot}, m_1)|s_1\rangle~
                \la L \frac{1}{2}; m s_1|L \frac{1}{2};\frac{1}{2} S_z\ra
                  ~\phi_{Lm}(x_1,x_2,x_3,k_{1\bot},k_{2\bot},k_{3\bot}),
\label{eq:Psi}
\end{equation}
where $\phi_{Lm}(x_1,x_2,x_3,k_{1\bot},k_{2\bot},k_{3\bot})$
describes the momentum distribution of the constituents in the
bound state with the subsystem consisting of the particles 2 and 3
in the orbital angular momentum $L,\,L_z=m$ state, $\la
L\frac{1}{2}; m s_1|1 \frac{1}{2};\frac{1}{2} S_z\ra$ is the
corresponding Clebsch-Gordan coefficient and $\langle
\lambda_1|{\cal R}_M^\dagger(x_1,k_{1\bot}, m_1)|s_1\rangle$ is
the well normalized Melosh transform matrix element. Explicitly
\cite{Jaus90,deAraujo:1999cr},
 \be
        \la \lambda_1|{\cal R}^\dagger_M (x_1,k_{1\bot},m_1)|s_1\ra
        &=&\frac{\bar
        u(k_1,\lambda) u_D(k_1,s_1)}{2 m_1}
        =\frac{(m_1+x_1 M_0)\delta_{\lambda_1 s_1}
                      -i\vec \sigma_{\lambda_1 s_1}\cdot\vec k_{1\bot} \times
                      \vec                n}
                {\sqrt{(m_1+x_1 M_0)^2 + k^{2}_{1\bot}}},
\en with $u_{(D)}$, a Dirac spinor in the light-front (instant)
form which has the expression
 \be
 u_D(k,s)=\frac{\not\!k+m}{\sqrt{k^0+m}}
 \left(\begin{array}{c}
       \chi_s\\
       0
       \end{array}
 \right),
 \quad
  u(k,\lambda)=\frac{\not\!k+m}{\sqrt{2 k^+}}\gamma^+\gamma^0
 \left(\begin{array}{c}
       \chi_\lambda\\
       0
       \end{array}
 \right),
 \label{eq:u}
 \en
in the Dirac representation, $\vec n = (0,0,1)$, a unit vector in
the $z$-direction, and
 \be
 M_0^2&=&\sum_{i=1}^3\frac{m_i^2+k^2_{i\bot}}{x_i},\quad
 k_i=(\frac{m_i^2+k^2_{i\bot}}{x_i M_0},x_i M_0,\,
 k_{i\bot})=(e_i-k_{iz},e_i+k_{iz},k_{i\bot}),
 \non\\
 M_0&=&e_1+e_2+e_3,\quad
 e_i =\sqrt{m^{2}_i+k^{2}_{i\bot}+k^{2}_{iz}}=\frac{x_i M_0}{2}+\frac{m_i^2+k^{2}_{i\bot}}{2 x_i M_0},\quad
 k_{iz}=\frac{x_i M_0}{2}-\frac{m_i^2+k^{2}_{i\bot}}{2 x_i M_0}.
 \non\\
 \en
Note that
 $u_D(k,s)=u(k,\lambda) \la \lambda|{\cal R}^\dagger_M|s\ra$
and, consequently, the state $|Q^c(k,\lambda)\ra \la \lambda|{\cal
R}^\dagger_M|s\ra$ transforms like $|Q^c(k,s)\ra$ under rotation,
i.e. its transformation does not depend on its momentum. A crucial
feature of the light-front formulation of a bound state, such as
the one shown in Eq.~(\ref{lfmbs}), is the frame-independence of
the light-front wave function~\cite{Brodsky:1997de,Jaus90}.
Namely, the hadron can be boosted to any (physical) ($P^+$,
$P_\bot$) without affecting the internal variables ($x_i$,
$k_{\bot i}$) of the wave function, which is certainly not the
case in the instant-form formulation.

In practice it is more convenient to use the covariant form for
the Melosh transform matrix element
\begin{equation}
       \langle \lambda_1|{\cal R}_M^\dagger(x_1,k_{1\bot}, m_1)|s_1\rangle~
                \la L \frac{1}{2}; m s_1|L \frac{1}{2};\frac{1}{2} S_z\ra
       \non\\=\frac{1}{\sqrt{2(p_1\cdot\bar P+m_1 M_0)}}
        ~\bar u(p_1,\lambda_1)\Gamma_{Lm} u(\bar P,S_z),
        \label{eq:covariant}
\end{equation}
with
\begin{eqnarray}
        &&\Gamma_{00}=1,\qquad
        \Gamma_{1m}=-\frac{1}{\sqrt3}\gamma_5\not\!\varepsilon^*(\bar
        P,m),\non\\
        &&\bar P\equiv p_1+p_2+p_3, \non \\
        &&\varepsilon^\mu(\bar P,\pm 1) =
                \left[{2\over P^+} \vec \varepsilon_\bot (\pm 1) \cdot
                \vec P_\bot,\,0,\,\vec \varepsilon_\bot (\pm 1)\right],
                \quad \vec \varepsilon_\bot
                (\pm 1)=\mp(1,\pm i)/\sqrt{2}, \nonumber\\
        &&\varepsilon^\mu(\bar P,0)={1\over M_0}\left({-M_0^2+P_\bot^2\over
                P^+},P^+,P_\bot\right),   \label{polcom}
\end{eqnarray}
for pentaquark states with $L=0$ or $L=1$ diquark pairs. To derive
the above expressions we have used the relations
 \be
 \bar u(k_1,\lambda_1)
 &=&\bar u(k_1,\lambda_1) \frac{u_D(k_1,s_1)\bar
 u_D(k_1,s_1)}{2 m_1}
 \non\\
 &=&\la\lambda_1|{\cal R}_M^\dagger(x_1,k_{1\bot}, m_1)|s_1\rangle\bar u_D(k_1,s_1),
 \non\\
 \la 0 \frac{1}{2}; 0 s_1|0 \frac{1}{2};\frac{1}{2} S_z\ra
 &=&\chi^\dagger_{s_1}\cdot\chi_{_{S_z}}
 \\
 &=&\frac{1}{\sqrt {2M_0 (e_1+m_1)}}\,
     \bar u_D(k_1,s_1) u(k_1+k_2+k_3,S_z),
 \non\\
 \la 1 \frac{1}{2}; m s_1|1 \frac{1}{2};\frac{1}{2} S_z\ra
 &=&-\frac{1}{\sqrt3}\chi^\dagger_{s_1}\vec\sigma\cdot
 \vec\varepsilon^*(k_1+k_2+k_3,m)\chi_{_{S_z}}
 \non\\
 &=&-\frac{1}{\sqrt {6 M_0 (e_1+m_1)}}\,
     \bar u_D(k_1,s_1)\gamma_5\not\!\varepsilon^*(k_1+k_2+k_3,m)
     u(k_1+k_2+k_3,S_z),
 \non
 \en
 where
$\chi_s$ is the usual Pauli spinor. The above relations can be
easily proved by using the explicit expression of the Dirac
spinors shown in Eq.~(\ref{eq:u}) and noting that
$k_1+k_2+k_3=(M_0,M_0,0_\bot)$. Putting these together and
boosting $k_i\to p_i$ we obtain Eq.~(\ref{eq:covariant}).
It should be remarked that in the conventional LF approach $\bar
P=p_1+p_2+p_3$ is not equal to the baryon's four-momentum  as all
constituents are on-shell and consequently $u(\bar P,S_z)$ is not
equal to $u(P,S_z)$; they satisfy different equations of motions
$(\not\!\!\bar P-M_0)u(\bar P,S_z)=0$ and $(\not\!\!P-M)u(
P,S_z)=0$. This is similar to the case of a vector meson bound
state where the polarization vectors $\varepsilon(\bar P,S_z)$ and
$\varepsilon(P,S_z)$ are different and satisfy different equations
$\varepsilon(\bar P,S_z)\cdot\bar P=0$ and
$\varepsilon(P,S_z)\cdot P=0$~\cite{Jaus91}. Although $u(\bar
P,S_z)$ is different than $u(P,S_z)$, they satisfy the relation
 \be
 \gamma^+ u(\bar P,S_z)=\gamma^+ u(P, S_z),
  \en
followed from $\gamma^+\gamma^+=0$, $\bar P^+=P^+$, $\bar
P_\bot=P_\bot$. This is again in analogy with the case of
$\varepsilon(\bar P,\pm 1)=\varepsilon(P,\pm 1)$.

The pentaquark baryon state is normalized as
\begin{equation}
        \langle \P(P',S'_z)|\P(P,S_z)\rangle = 2(2\pi)^3 P^+
        \delta^3(\tilde P'- \tilde P)\delta_{L'L}\delta_{S'_z S_z}~,
\label{wavenor}
\end{equation}
so that [cf. Eqs. (\ref{lfmbs}), (\ref{eq:norm}) and
(\ref{eq:Psi})]
\begin{equation}
        \int \left(\Pi_{i=1}^3{dx_i\,d^2k_{i\bot}\over 2(2\pi)^3}\right)
        2(2\pi)^3\delta(1-\sum x_i)\delta^2(\sum k_{i\bot})
        ~\phi^{\prime*}_{Lm}(\{x\},\{k_\bot\}) \phi_{Lm}(\{x\},\{k_\bot\})
        =~\delta_{L'L}\delta_{m'm}.
\label{momnor}
\end{equation}
Under the constraint of $1-\sum_{i=1}^3 x_i=\sum_{i=1}^3
(k_i)_{x,y,z}=0$, we have the expressions
 \be
   &&\phi_{Lm}(\{x\},\{k_\bot\})=\sqrt{\frac{\partial(k_{2z}, k_{3z})}{\partial
   (x_2,
  x_3)}}\,\varphi_{00}(\vec k_1,\beta_1)~\varphi_{Lm}
  \left(\frac{\vec k_2-\vec k_3}{2},\beta_{23}\right), \non\\
  &&\frac{\partial(k_{2z}, k_{3z})}{\partial (x_2,
  x_3)}=\frac{e_1 e_2 e_3}{x_1 x_2 x_3 M_0},\quad
  \varphi_{00}(\vec k,\beta)=\varphi(\vec k,\beta),\quad
  \varphi_{1m}(\vec k,\beta)=k_{m} \varphi_p(\vec k,\beta),
  \label{eq:phi}
 \en
where $k_m=\vec\varepsilon(m)\cdot\vec k$, or explicitly
$k_{m=\pm1}=\mp(k_{\bot x}\pm i k_{\bot y})/\sqrt2$,
$k_{m=0}=k_{z}$, are proportional to the spherical harmonics
$Y_{1m}$ in the momentum space, and $\varphi$, $\varphi_p$ are the
distribution amplitudes of $s$-wave and $p$-wave states,
respectively. For a Gaussian-like wave function, one has
\cite{Cheng97,CCH}
\begin{eqnarray} \label{eq:Gauss}
 \varphi(\vec k,\beta)
    &=&4 \left({\pi\over{\beta^{2}}}\right)^{3\over{4}}
               ~{\rm exp}
               \left(-{k^2_z+k^2_\bot\over{2
               \beta^2}}\right),\quad
    \varphi_p(\vec k,\beta)=\sqrt{2\over{\beta^2}}~\varphi(\vec k,\beta).
 \label{eq:wavefn}
\end{eqnarray}
In order to see that the above wave functions do satisfy the
normalization condition (\ref{momnor}), we note that (under the
above-mentioned constraints)
 \be
 dx_2 d^2 k_{2\bot}~dx_3 d^2 k_{3\bot}
 \frac{\partial(k_{2z}, k_{3z})}{\partial (x_2, x_3)}
 =d^3k_2~d^3 k_3
 = d^3(k_2+k_3)~d^3\left(\frac{k_2-k_3}{2}\right),
 \en
with $d^3(k_2+k_3)=d^3 k_1$ and the Gaussian-like wave functions
can be integrated readily. It is easily seen that $\phi_{Lm}$
picks up a factor of $(-)^L$ under the interchange of
$k_2\leftrightarrow k_3$. This will compensate the $(-)^L$ factor
appearing in Eq.~(\ref{eq:norm}).

By using Eq.~(\ref{eq:covariant}) it is straightforward to obtain
 \be
 &&\langle \lambda_1|{\cal R}_M^\dagger(x_1,k_{1\bot}, m_1)|s_1\rangle~
                \la 1 \frac{1}{2}; m s_1|1 \frac{1}{2};\frac{1}{2} S_z\ra \frac{(k_2-k_3)_m}{2}
 \non\\
 &&=\frac{1}{2\sqrt{6(p_1\cdot\bar P+m_1 M_0)}}
        ~\bar u(p_1,\lambda_1)\gamma_5\left[\not\!p_2-\not\!p_3-\frac{\bar P\cdot (p_2-p_3)}{M_0}\right]u(\bar
        P,S_z).
         \label{eq:covariantP}
 \en
where the factor of $(k_2-k_3)_m=\varepsilon(\bar
P,m)\cdot(p_2-p_3)$ comes from the wave function
Eq.~(\ref{eq:phi}) for the $L=1$ case. The state $|\P(P,L,
S_z)\rangle$ for $\P_Q$ in the light-front model can now be
obtained by using Eqs.~(\ref{lfmbs})--(\ref{eq:covariantP}).

\subsection{$\P_Q\to \P_{Q'}$ weak transitions}

In this work we consider $\P_{b}\to \P_{c}$ and $\P_{c}\to \P_{s}$
(especially $\Theta_{c}\to \Theta_{s}$) weak transitions as
depicted in Fig.~\ref{fig:penta}. The matrix element for the
$\P_{Q'}\to \P_{Q}$ weak transition is
 \be
 \la \P_{Q'}(P',S'_z)|\bar Q\gamma_\mu (1-\gamma_5)
 Q'|\P_{Q}(P,S_z)\ra
 =\la \P_{Q'}(P',S'_z)|-\bar Q^{\prime c}\gamma_\mu (1+\gamma_5)
 Q^c|\P_{Q}(P,S_z)\ra,
 \label{eq:antiQ}
 \en
where $Q^{\prime c}$ and $Q^c$ are the charge conjugated fields of
$Q'$ and $Q$, respectively. The reason we use the charge
conjugated field for $\bar Q$ in the above equation and in the
previous subsection is to have both $Q^c$ and $Q'_c$ fermion lines
flow in the same direction as the flow of the pentaquark fermion
line such that we do not need a further transpose of Dirac
matrices in the ensuing calculation. If we treat $\bar Q$ as an
anti-fermion we will have to use $v^T\Gamma' u$ instead of $\bar u
\Gamma u$ in Eq.~(\ref{eq:covariant}). The above matrix element
can be parameterized as
 \be
 &&\la \P_{Q'}(P',S'_z)|\bar Q\gamma_\mu  Q'|\P_Q(P,S_z)\ra
 \non\\
 &&\qquad=-\bar u(P',S'_z)\Big[f_1(q^2)\gamma_\mu+i{f_2(q^2)\over M+M'} \sigma_{\mu\nu}q^\nu
    +{f_3(q^2)\over M+M'}q_\mu\Big] u(P,S_z),
 \non\\
 &&\la \P_{Q'}(P',S'_z)|\bar Q\gamma_\mu \gamma_5 Q'|\P_Q(P,S_z)\ra
 \non\\
 &&\qquad=\bar u(P',S'_z)\Big[g_1(q^2)\gamma_\mu+i{g_2(q^2)\over M+M'} \sigma_{\mu\nu}q^\nu
     +{g_3(q^2)\over M+M'}q_\mu\Big]\gamma_5 u(P,S_z),
 \label{eq:figi}
 \en
with $q=P-P'$. Note that an overall $``-"$ sign arises in the
first matrix element owing to a minus sign in front of $\bar
Q^{\prime c}\gamma_\mu Q^c$ in Eq.~(\ref{eq:antiQ}).

Armed with the light-front quark model description of
$|\P_Q(P,S_z)\ra$ in the previous subsection, we are ready to
calculate the weak transition matrix element of heavy pentaquarks.
For $\P_Q({\bf 3})\to\P_{Q'}({\bf 3})$ ($L=0$) and $\P_Q(\bar {\bf
6})\to\P_{Q'}(\bar{\bf 6})$ ($L=1$) transitions, we have the
general expressions (after using Eqs.~(\ref{lfmbs}),
(\ref{eq:antiQ}) and integrating out the heavy quark momentum
$p_1$)
 \be
 &&\la \P_{Q'}(P',S'_z)|\bar Q\gamma^\mu (1-\gamma_5)
 Q'|\P_{Q}(P,S_z)\ra
 \non\\
 &&=\int \{d^3p_2\}\{d^3p_3\}~
 \frac{\phi^{\prime*}_{Lm'}(\{x'\},\{k'_{\bot}\})\phi_{Lm}(\{x\},\{k_{\bot}\})}
 {2\sqrt{p_1^+ p_1^{\prime +}P^+ P^{\prime +}(p_1\cdot \bar P+m_1 M_0)(p'_1\cdot\bar P'+m'_1 M'_0)}}
 \non\\
 &&\qquad\qquad\qquad\qquad\times~\bar u(\bar P',S'_z)\bar \Gamma^\prime_{Lm} (\not\!
 p'_1+m'_1)\gamma^\mu(-1-\gamma_5)(\not\!
 p_1+m_1)\Gamma_{Lm} u(\bar P,S_z),
 \label{eq:P->P'}
 \en
where the diquark pairs act as spectators, $\bar
\Gamma_{Lm}=\gamma_0\Gamma^\dagger_{Lm} \gamma_0$ and
\begin{eqnarray}
        && p^{(\prime)+}_i=x^{(\prime)}_i P^{(\prime)+},\qquad
           p^{(\prime)}_{i\bot}=x^{(\prime)}_i
           P^{(\prime)}_\bot+k^{(\prime)}_{i\bot},\qquad 1-\sum_{i=1}^3
           x^{(\prime)}_i=\sum_{i=1}^3 k^{(\prime)}_{i\bot}=0,
        \non\\
        && \tilde p_1-\tilde p_1^{\prime}=\tilde q,\qquad
        \tilde p_2=\tilde p_2^{\prime},\qquad
        \tilde p_3=\tilde p_3^{\prime},
\end{eqnarray}
with $\tilde p=(p^+,\,p_\bot)$ and $\Gamma_{Lm}$ given in Eq.
(\ref{polcom}). Explicit expressions for $\P_Q({\bf 3})\to
\P_{Q'}({\bf 3})$ and $\P_Q({\6bar})\to \P_{Q'}({\6bar})$
transition matrix elements will be given later.

We shall follow~\cite{Schlumpf} to project out various form
factors from the transition matrix elements. As in
\cite{CCH,Schlumpf}, we consider the $q^+=0$, $q_\bot\not=0$ case.
To proceed, we apply the relations
 \be
 \frac{\bar u(P',S'_z)\gamma^+ u(P, S_z)}{2\sqrt{P^+ P^{\prime
 +}}}&=&\delta_{S'_z S_z},
 \qquad
 i\frac{\bar u(P',S'_z)\sigma^{+\nu} q_\nu u(P, S_z)}{2\sqrt{P^+
P^{\prime
 +}}}=(\vec\sigma\cdot \vec q_\bot\sigma^3)_{S'_z S_z},
  \non\\
  \frac{\bar u(P',S'_z)\gamma^+\gamma_5 u(P, S_z)}{2\sqrt{P^+ P^{\prime
 +}}}&=&(\sigma^3)_{S'_z S_z},
 \quad
 i\frac{\bar u(P',S'_z)\sigma^{+\nu} q_\nu\gamma_5 u(P, S_z)}{2\sqrt{P^+
P^{\prime
 +}}}=(\vec\sigma\cdot \vec q_\bot)_{S'_z S_z},
 \label{eq:spinorprojection}
  \en
extended from the first identity (as shown in \cite{Pauli}) by
applying Eq.~(\ref{eq:u}), and obtain~\cite{Schlumpf}
 \be
 f_1(q^2)
 &=&-\frac{\la \P_{Q'}(P',\uparrow)|V^+|\P_Q(P,\uparrow)\ra}{2 \sqrt{P^+ P^{\prime +}}}
         =-\frac{\la \P_{Q'}(P',\downarrow)|V^+|\P_Q(P,\downarrow)\ra}{2\sqrt{P^+ P^{\prime+}}},
 \non\\
 \frac{f_2(q^2)}{M+M'}
 &=&\frac{\la \P_{Q'}(P',\uparrow)|V^+|\P_Q(P,\downarrow)\ra}{2q_{\bot L}\sqrt{P^+ P^{\prime+}}}
         =-\frac{\la \P_{Q'}(P',\downarrow)|V^+|\P_Q(P,\uparrow)\ra}{2q_{\bot R}\sqrt{P^+ P^{\prime+}}},
 \non\\
 g_1(q^2)
 &=&\frac{\la \P_{Q'}(P',\uparrow)|A^+|\P_Q(P,\uparrow)\ra}{2 \sqrt{P^+ P^{\prime+}}}
         =-\frac{\la \P_{Q'}(P',\downarrow)|A^+|\P_Q(P,\downarrow)\ra}{2\sqrt{P^+ P^{\prime +}}},
 \non\\
 \frac{g_2(q^2)}{M+M'}
 &=&\frac{\la \P_{Q'}(P',\uparrow)|A^+|\P_Q(P,\downarrow)\ra}{2q_{\bot L}\sqrt{P^+ P^{\prime+}}}
         =\frac{\la \P_{Q'}(P',\downarrow)|A^+|\P_Q(P,\uparrow)\ra}{2q_{\bot R}\sqrt{P^+ P^{\prime+}}},
 \en
where $q_{\bot L,R}=q_\bot^1\mp i q_\bot^2$, or equivalently, we
have
 \be
 \la \P_{Q'}(P',S'_z)|V^+|\P_Q(P,S_z)\ra
 &=&-2\sqrt{P^+ P^{\prime +}}
 \left[f_1(q^2)~\delta_{S'_z S_z}+\frac{f_2(q^2)}{M+M'}(\vec\sigma\cdot \vec q_\bot\sigma^3)_{S'_z
 S_z}\right],
 \non\\
 \la \P_{Q'}(P',S'_z)|A^+|\P_Q(P,S_z)\ra
 &=&2\sqrt{P^+P^{\prime +}}
 \left[ g_1(q^2)~(\sigma^3)_{S'_z S_z}+\frac{g_2(q^2)}{M+M'}(\vec\sigma\cdot \vec q_\bot)_{S'_z
 S_z}\right].
 \label{eq:projection}
  \en
Various form factors can be projected out by applying the
orthogonality of the corresponding matrices, $\delta_{S'_z S_z}$,
$(\sigma_{\bot}^i\sigma^3)_{S'_z S_z}$, $(\sigma^3)_{S'_z S_z}$
and $(\sigma_\bot^i)_{S'_z S_z}$, under the trace operation. Note
that due to the condition $q^+=0$ we have imposed in passing, the
form factors $f_3(q^3)$ and $g_3(q^3)$ cannot be extracted in this
manner. To extract $f_{1,2}(q^2)$ and $g_{1,2}(q^2)$ from the
right hand side of Eq.~(\ref{eq:P->P'}), the following identities
are proved to be useful:
  \be
 \frac{1}{2}\sum_{S_z,S'_z} u(\bar P, S_z)\delta_{S_z S'_z}\bar u(\bar P',S'_z)
  &=&\frac{1}{4\sqrt{P^+P^{\prime+}}}(\not\!\bar P+M_0)\gamma^+(\not\!\bar P'+M'_0),
 \non\\
 \frac{1}{2}\sum_{S_z,S'_z} u(\bar P, S_z)(\sigma^3\sigma_{\bot}^i)_{S_z S'_z}\bar u(\bar P',S'_z)
  &=&-\frac{i}{4\sqrt{P^+P^{\prime+}}}(\not\!\bar P+M_0)\sigma^{i+}(\not\!\bar P'+M'_0),
 \non\\
 \frac{1}{2}\sum_{S_z,S'_z} u(\bar P, S_z)(\sigma^3)_{S_z S'_z}\bar u(\bar P',S'_z)
  &=&\frac{1}{4\sqrt{P^+P^{\prime+}}}(\not\!\bar P+M_0)\gamma^+\gamma_5(\not\!\bar P'+M'_0),
 \non\\
 \frac{1}{2}\sum_{S_z,S'_z} u(\bar P, S_z)(\sigma_\bot^i)_{S_z S'_z}\bar u(\bar P',S'_z)
  &=&\frac{i}{4\sqrt{P^+P^{\prime+}}}(\not\!\bar P+M_0)\sigma^{i+}\gamma_5(\not\!\bar
  P'+M'_0).
  \label{eq:spinorprojectionbar}
 \en
With the above generic discussions on $\P_Q\to \P_{Q'}$
transition, we are ready to extract $\P_Q({\bf 3})\to \P_{Q'}({\bf
3})$ and $\P_Q(\6bar)\to \P_{Q'}(\6bar)$ transition form factors.

\subsubsection{Form factors for the $\P_Q({\bf 3})\to\P_{Q'}({\bf 3})$
transition}

Following from Eq.~(\ref{eq:P->P'}), we have explicitly
 \be
 &&\la \P_{Q'}(P',S'_z)|\bar Q\gamma^\mu (1-\gamma_5)
 Q'|\P_{Q}(P,S_z)\ra
 \non\\
 &&=\int \frac{dx_2 d^2 k_{2\bot}}{2 (2\pi)^3}\frac{dx_3 d^2 k_{3\bot}}{2 (2\pi)^3}~
 \frac{\phi^{\prime*}_{00}(\{x'\},\{k'_{\bot}\})~\phi_{00}(\{x\},\{k_{\bot}\})}
 {2\sqrt{x_1 x'_1(p_1\cdot \bar P+m_1 M_0)(p'_1\cdot\bar P'+m'_1 M'_0)}}
 \non\\
 &&\qquad\qquad\qquad\qquad\times~\bar u(\bar P',S'_z) (\not\!
 p'_1+m'_1)\gamma^\mu(-1-\gamma_5)(\not\!
 p_1+m_1) u(\bar P,S_z),
 \label{eq:3->3}
 \en
for the $\P_Q({\bf 3})\to\P_{Q'}({\bf 3})$ transition. By using
Eqs. (\ref{eq:3->3}), (\ref{eq:projection}) and
(\ref{eq:spinorprojectionbar})
we obtain the $\P_Q({\bf 3})\to \P_{Q'}({\bf 3})$ transition form
factors
 \be
 f_1(q^2)&=&\frac{1}{8 P^+P^{\prime+}}\int \frac{dx_2 d^2 k_{2\bot}}{2 (2\pi)^3}\frac{dx_3 d^2 k_{3\bot}}{2 (2\pi)^3}~
 \frac{\phi^{\prime*}_{00}(\{x'\},\{k'_{\bot}\})~\phi_{00}(\{x\},\{k_{\bot}\})}
 {2\sqrt{x_1 x'_1(p_1\cdot \bar P+m_1 M_0)(p'_1\cdot\bar P'+m'_1 M'_0)}}
 \non\\
 &&\qquad\qquad\times~{\rm Tr}[(\not\!\bar P+M_0)\gamma^+(\not\!\bar P'+M'_0) (\not\!
 p'_1+m'_1)\gamma^+(\not\!
 p_1+m_1)],
 \non\\
 \frac{f_2(q^2)}{M+M'}&=&-\frac{i}{8 P^+P^{\prime+} q_\bot^i}\int \frac{dx_2 d^2 k_{2\bot}}{2 (2\pi)^3}\frac{dx_3 d^2 k_{3\bot}}{2 (2\pi)^3}~
 \frac{\phi^{\prime*}_{00}(\{x'\},\{k'_{\bot}\})~\phi_{00}(\{x\},\{k_{\bot}\})}
 {2\sqrt{x_1 x'_1(p_1\cdot \bar P+m_1 M_0)(p'_1\cdot\bar P'+m'_1 M'_0)}}
 \non\\
 &&\qquad\qquad\times~{\rm Tr}[(\not\!\bar P+M_0)\sigma^{i+}(\not\!\bar P'+M'_0) (\not\!
 p'_1+m'_1)\gamma^+(\not\!
 p_1+m_1)],
 \non\\
  g_1(q^2)&=&\frac{1}{8 P^+P^{\prime+}}\int \frac{dx_2 d^2 k_{2\bot}}{2 (2\pi)^3}\frac{dx_3 d^2 k_{3\bot}}{2 (2\pi)^3}~
 \frac{\phi^{\prime*}_{00}(\{x'\},\{k'_{\bot}\})~\phi_{00}(\{x\},\{k_{\bot}\})}
 {2\sqrt{x_1 x'_1(p_1\cdot \bar P+m_1 M_0)(p'_1\cdot\bar P'+m'_1 M'_0)}}
 \non\\
 &&\qquad\qquad\times~{\rm Tr}[(\not\!\bar P+M_0)\gamma^+\gamma_5(\not\!\bar P'+M'_0) (\not\!
 p'_1+m'_1)\gamma^+\gamma_5(\not\!
 p_1+m_1)],
 \non\\
 \frac{g_2(q^2)}{M+M'}&=&\frac{i}{8 P^+P^{\prime+} q_\bot^i}\int \frac{dx_2 d^2 k_{2\bot}}{2 (2\pi)^3}\frac{dx_3 d^2 k_{3\bot}}{2 (2\pi)^3}~
 \frac{\phi^{\prime*}_{00}(\{x'\},\{k'_{\bot}\})~\phi_{00}(\{x\},\{k_{\bot}\})}
 {2\sqrt{x_1 x'_1(p_1\cdot \bar P+m_1 M_0)(p'_1\cdot\bar P'+m'_1 M'_0)}}
 \non\\
 &&\qquad\qquad\times~{\rm Tr}[(\not\!\bar P+M_0)\sigma^{i+}\gamma_5(\not\!\bar P'+M'_0) (\not\!
 p'_1+m'_1)\gamma^+\gamma_5(\not\!
 p_1+m_1)],
 \label{eq:ff3->3compact}
 \en
with $q_\bot^ i=q_\bot^1$ or $q_\bot^2$ (no sum over $i$).

It is straightforward to work out the traces in $f_{1,2}(q^2)$ as
shown in Eq.~(\ref{eq:ff3->3compact}) and obtain
 \be
 &&\frac{1}{8 P^+ P^{\prime+}}{\rm Tr}[(\not\!\bar P+M_0)\gamma^+(\not\!\bar P'+M'_0) (\not\!
 p'_1+m'_1)\gamma^+(\not\! p_1+m_1)]
 \non\\
 &&\qquad\qquad=-(p_1-x_1\bar P)\cdot (p'_1-x'_1 \bar P')+(x_1 M_0+m_1)(x'_1
 M'_0+m'_1),
 \non\\
&&\frac{i}{8 P^+ P^{\prime+}}{\rm Tr}[(\not\!\bar
P+M_0)\sigma^{i+}(\not\!\bar P'+M'_0) (\not\!
 p'_1+m'_1)\gamma^+(\not\!
 p_1+m_1)]
 \non\\
 &&\qquad\qquad=(m'_1+x'_1 M'_0) (p^i_\bot-x_1 \bar P^i_\bot)
 -(m_1+x_1 M_0) (p^{\prime i}_\bot-x'_1 \bar P^{\prime i}_\bot),
 \label{eq:trace1}
 \en
for $i=1,2$, where uses of  $\bar P^{(\prime)+}=P^{(\prime)+}$,
$\bar P_\bot^{(\prime)i}=P^{(\prime)i}_\bot$, $p^{(\prime)
+}_1=x^{(\prime)} P^{(\prime)+}$, $p^{(\prime)
i}_{1\bot}=x^{(\prime)} P^{(\prime)i}_\bot+k^{(\prime)i}_{1\bot}$
have been made. Note that the traces in $g_1(q^2)$ and $g_2(q^2)$
under the replacement $m'_1\to-m'_1$, $M'_0\to -M'_0$ are the same
as that in $f_1(q^2)$ and $f_2(q^2)$, respectively, except for an
additional overall minus sign. Then the traces and the common
dominator factors in the above form factors can be expressed in
terms of the internal variables via
 \be
 &&p_1\cdot \bar P=e_1 M_0=\frac{m_1^2+x^2_1 M_0^2+k^2_{1\bot}}{2 x_1},\quad
 p'_1\cdot \bar P'=e'_1 M'_0=\frac{m_1^{\prime 2}+x^2_1 M_0^{\prime 2}+k^{\prime 2}_{1\bot}}{2 x_1},
 \non\\
 &&(p_1-x_1\bar P)\cdot (p'_1-x'_1 \bar P')=-k_{1\bot}\cdot
 k'_{1\bot},\quad
 p^{(\prime)i}_\bot-x_1 \bar
 P^{(\prime)i}_\bot=k_{1\bot}^{(\prime)i},
 \label{eq:internal1}
 \en
where $k_{1\bot}\cdot
 k'_{1\bot}$ is a scalar product in two-dimensional space.
Using Eqs.~(\ref{eq:ff3->3compact}), (\ref{eq:trace1}) and
(\ref{eq:internal1}) we obtain the explicit forms of the
$\P_Q({\bf 3})\to \P_{Q'}({\bf 3})$ transition form factors
 \be
 f_1(q^2)&=&\int \frac{dx_2 d^2 k_{2\bot}}{2 (2\pi)^3}\frac{dx_3 d^2 k_{3\bot}}{2 (2\pi)^3}~
 \frac{\phi^{\prime*}_{00}(\{x'\},\{k'_{\bot}\})~\phi_{00}(\{x\},\{k_{\bot}\})}
 {\sqrt{[(m_1+x_1 M_0)^2+k_{1\bot}^2][(m'_1+x_1 M'_0)^2+k_{1\bot}^{\prime 2}]}}
 \non\\
 &&\qquad\qquad\qquad\qquad\qquad\times~
   [k_{1\bot}\cdot k'_{1\bot}+(m_1+x_1 M_0)(m'_1+x'_1 M'_0)],
 \non\\
 \frac{f_2(q^2)}{M+M'}&=&\frac{1}{q^i_\bot}\int \frac{dx_2 d^2 k_{2\bot}}{2 (2\pi)^3}\frac{dx_3 d^2 k_{3\bot}}{2 (2\pi)^3}~
 \frac{\phi^{\prime*}_{00}(\{x'\},\{k'_{\bot}\})~\phi_{00}(\{x\},\{k_{\bot}\})}
 {\sqrt{[(m_1+x_1 M_0)^2+k_{1\bot}^2][(m'_1+x_1 M'_0)^2+k_{1\bot}^{\prime 2}]}}
 \non\\
 &&\qquad\qquad\qquad\qquad\qquad\times~
   [(m_1+x_1 M_0)~k^{\prime i}_{1\bot}-(m'_1+x'_1 M'_0)~k^i_{1\bot}],
 \non\\
  g_1(q^2)&=&\int \frac{dx_2 d^2 k_{2\bot}}{2 (2\pi)^3}\frac{dx_3 d^2 k_{3\bot}}{2 (2\pi)^3}~
 \frac{\phi^{\prime*}_{00}(\{x'\},\{k'_{\bot}\})~\phi_{00}(\{x\},\{k_{\bot}\})}
 {\sqrt{[(m_1+x_1 M_0)^2+k_{1\bot}^2][(m'_1+x_1 M'_0)^2+k_{1\bot}^{\prime 2}]}}
 \non\\
 &&\qquad\qquad\qquad\qquad\qquad\times~
   [-k_{1\bot}\cdot k'_{1\bot}+(m_1+x_1 M_0)(m'_1+x'_1 M'_0)],
 \non\\
 \frac{g_2(q^2)}{M+M'}&=&\frac{1}{q^i_\bot}\int \frac{dx_2 d^2 k_{2\bot}}{2 (2\pi)^3}\frac{dx_3 d^2 k_{3\bot}}{2 (2\pi)^3}~
 \frac{\phi^{\prime*}_{00}(\{x'\},\{k'_{\bot}\})~\phi_{00}(\{x\},\{k_{\bot}\})}
 {\sqrt{[(m_1+x_1 M_0)^2+k_{1\bot}^2][(m'_1+x_1 M'_0)^2+k_{1\bot}^{\prime 2}]}}
 \non\\
 &&\qquad\qquad\qquad\qquad\qquad\times~
   [(m'_1+x'_1 M'_0)~k^i_{1\bot}+(m_1+x_1 M_0)~k^{\prime i}_{1\bot}],
 \label{eq:ff3->3}
 \en
with $q_\bot^ i=q_\bot^1$ or $q_\bot^2$ (no sum over $i$).

\subsubsection{Form factors for the $\P_Q({\6bar})\to\P_{Q'}({\6bar})$
transition}

For the $\P_Q(\bar {\bf 6})\to\P_{Q'}(\bar{\bf 6})$ transition,
Eq.~(\ref{eq:P->P'}) leads to
 \be
 &&\la \P_{Q'}(P',S'_z)|\bar Q\gamma^\mu (1-\gamma_5)
 Q'|\P_{Q}(P,S_z)\ra
 \non\\
 &&=\int \frac{dx_2 d^2 k_{2\bot}}{2 (2\pi)^3}\frac{dx_3 d^2 k_{3\bot}}{2 (2\pi)^3}~
 \frac{\phi^{\prime*}_{00}(\{x'\},\{k'_{\bot}\})~\phi_{00}(\{x\},\{k_{\bot}\})}
 {12\beta_{23}\beta'_{23}\sqrt{x_1 x'_1(p_1\cdot \bar P+m_1 M_0)(p'_1\cdot\bar P'+m'_1 M'_0)}}
 ~(p'_2-p'_3)_\rho(p_2-p_3)_\sigma
 \non\\
 &&~\times~ \bar u(\bar P',S'_z) \gamma_5 \Big(\gamma^\rho+\frac{\bar P^{\prime\rho}}{M'_0}\Big)(\not\!
 p'_1+m'_1)\gamma^\mu(-1-\gamma_5)(\not\!
 p_1+m_1) \gamma_5\Big(\gamma^\sigma-\frac{\bar P^{\sigma}}{M_0}\Big) u(\bar P,S_z).
 \label{eq:6->6}
 \en
By repeating the same procedure of the form factor extraction as
before, we obtain the $\P_Q({\6bar})\to \P_{Q'}({\6bar})$
transition form factors
 \be
 f_1(q^2)&=&\frac{1}{8 P^+P^{\prime+}}\int \frac{dx_2 d^2 k_{2\bot}}{2 (2\pi)^3}\frac{dx_3 d^2 k_{3\bot}}{2 (2\pi)^3}~
 \frac{\phi^{\prime*}_{00}(\{x'\},\{k'_{\bot}\})~\phi_{00}(\{x\},\{k_{\bot}\})~(p'_2-p'_3)_\rho(p_2-p_3)_\sigma}
 {12\beta_{23}\beta'_{23}\sqrt{x_1 x'_1(p_1\cdot \bar P+m_1 M_0)(p'_1\cdot\bar P'+m'_1 M'_0)}}
 \non\\
 &\times&{\rm Tr}\Big[(\not\!\bar P+M_0)\gamma^+(\not\!\bar P'+M'_0)
  \Big(\gamma^\rho-\frac{\bar P^{\prime\rho}}{M'_0}\Big)(\not\!
 p'_1-m'_1)\gamma^+(\not\!
 p_1-m_1)\Big(\gamma^\sigma-\frac{\bar P^{\sigma}}{M_0}\Big)\Big],
 \non\\
 \frac{f_2(q^2)}{M+M'}&=&-\frac{i}{8 P^+P^{\prime+} q_\bot^i}\int \frac{dx_2 d^2 k_{2\bot}}{2 (2\pi)^3}\frac{dx_3 d^2 k_{3\bot}}{2 (2\pi)^3}~
 \frac{\phi^{\prime*}_{00}(\{x'\},\{k'_{\bot}\})~\phi_{00}(\{x\},\{k_{\bot}\})~(p'_2-p'_3)_\rho(p_2-p_3)_\sigma}
 {12\beta_{23}\beta'_{23}\sqrt{x_1 x'_1(p_1\cdot \bar P+m_1 M_0)(p'_1\cdot\bar P'+m'_1 M'_0)}}
 \non\\
 &\times&{\rm Tr}\Big[(\not\!\bar P+M_0)\sigma^{i+}(\not\!\bar P'+M'_0)
  \Big(\gamma^\rho-\frac{\bar P^{\prime\rho}}{M'_0}\Big)(\not\!
 p'_1-m'_1)\gamma^+(\not\!
 p_1-m_1)\Big(\gamma^\sigma-\frac{\bar P^{\sigma}}{M_0}\Big)\Big],
 \non\\
 g_1(q^2)&=&\frac{1}{8 P^+P^{\prime+}}\int \frac{dx_2 d^2 k_{2\bot}}{2 (2\pi)^3}\frac{dx_3 d^2 k_{3\bot}}{2 (2\pi)^3}~
 \frac{\phi^{\prime*}_{00}(\{x'\},\{k'_{\bot}\})~\phi_{00}(\{x\},\{k_{\bot}\})~(p'_2-p'_3)_\rho(p_2-p_3)_\sigma}
 {12\beta_{23}\beta'_{23}\sqrt{x_1 x'_1(p_1\cdot \bar P+m_1 M_0)(p'_1\cdot\bar P'+m'_1 M'_0)}}
 \non\\
 &\times&{\rm Tr}\Big[(\not\!\bar P+M_0)\gamma^+\gamma_5(\not\!\bar P'+M'_0)
  \Big(\gamma^\rho-\frac{\bar P^{\prime\rho}}{M'_0}\Big)(\not\!
 p'_1-m'_1)\gamma^+\gamma_5(\not\!
 p_1-m_1)\Big(\gamma^\sigma-\frac{\bar P^{\sigma}}{M_0}\Big)\Big],
 \non\\
 \frac{g_2(q^2)}{M+M'}&=&\frac{i}{8 P^+P^{\prime+} q_\bot^i}\int \frac{dx_2 d^2 k_{2\bot}}{2 (2\pi)^3}\frac{dx_3 d^2 k_{3\bot}}{2 (2\pi)^3}~
 \frac{\phi^{\prime*}_{00}(\{x'\},\{k'_{\bot}\})~\phi_{00}(\{x\},\{k_{\bot}\})~(p'_2-p'_3)_\rho(p_2-p_3)_\sigma}
 {12\beta_{23}\beta'_{23}\sqrt{x_1 x'_1(p_1\cdot \bar P+m_1 M_0)(p'_1\cdot\bar P'+m'_1 M'_0)}}
 \non\\
 &\times&{\rm Tr}\Big[(\not\!\bar P+M_0)\sigma^{i+}\gamma_5(\not\!\bar P'+M'_0)
  \Big(\gamma^\rho-\frac{\bar P^{\prime\rho}}{M'_0}\Big)(\not\!
 p'_1-m'_1)\gamma^+\gamma_5(\not\!
 p_1-m_1)\Big(\gamma^\sigma-\frac{\bar P^{\sigma}}{M_0}\Big)\Big],
 \label{eq:ff6->6compact}
 \en
with $q_\bot^ i=q_\bot^1$ or $q_\bot^2$ (no sum over $i$).

For the traces in $f_{1,2}(q^2)$ in Eq.~(\ref{eq:ff6->6compact}),
we have
 \be
 &&\frac{1}{8 P^+ P^{\prime +}}{\rm Tr}\Big[(\not\!\bar P+M_0)\gamma^+(\not\!\bar P'+M'_0)
  \Big(\not\!p_{23}-\frac{\bar P'\cdot p_{23}}{M'_0}\Big)(\not\!  p'_1-m'_1)\gamma^+
  (\not\! p_1-m_1)\Big(\not\!p_{23}-\frac{\bar P\cdot p_{23}}{M_0}\Big)\Big]
 \non\\
 &&\quad=p^2_{23}[(p_1-x_1 \bar P)\cdot(p'_1-x_1 \bar P')
             -(m_1+x_1 M_0)(m'_1+x_1 M'_0)]
 \non\\
 &&\quad     +x^2_{23}[(p_1\cdot \bar P+m_1 M_0) (p'_1\cdot\bar P'+m'_1 M'_0)
             +(p_1\cdot p'_1-m_1 m'_1)(\bar P\cdot\bar P'-M_0 M'_0)
 \non\\
 &&\quad     -(p_1\cdot\bar P'+m_1 M'_0) (p'_1\cdot\bar P+m'_1 M_0)]
        +\frac{p_1\cdot p_{23}}{M_0}
             \{-(1-x_1) p_1\cdot p_{23} (m'_1+x_1 M'_0)
 \non\\
 &&\quad     +x_{23}[(p_1\cdot\bar P'+m_1 M'_0)(m'_1+x_1 M_0)
             +(p'_1\cdot\bar P+m'_1 M_0)(M_0-x_1 M'_0)
 \non\\
 &&\quad-(p'_1\cdot\bar P'+m'_1 M'_0)(m_1+M_0)
             +(p_1\cdot p'_1-m_1 m'_1)(M'_0-M_0)
             -x_1(\bar P\cdot\bar P'-M_0 M'_0)(m'_1+M_0)]\}
 \non\\
 &&\quad+\frac{p'_1\cdot p_{23}}{M'_0}
             \{-(1-x_1) p'_1\cdot p_{23} (m_1+x_1 M_0)
             +x_{23}[(p'_1\cdot\bar P+m'_1 M_0)(m_1+x_1 M'_0)
 \non\\
 &&\quad     +(p_1\cdot\bar P'+m_1 M'_0)(M'_0-x_1 M_0)
             -(p_1\cdot\bar P+m_1 M_0)(m'_1+M'_0)
 \non\\
 &&\quad     +(p_1\cdot p'_1-m_1 m'_1)(M_0-M'_0)
             -x_1(\bar P\cdot\bar P'-M_0 M'_0)(m_1+M'_0)]\}
 \non\\
 &&\quad+\frac{p_1\cdot p_{23}~p'_1\cdot p_{23}}{M_0 M'_0}
             [-(p_1-x_1\bar P)\cdot(p'_1-x_1\bar P')+(m_1+M_0)(m'_1+M'_0)-M_0 M'_0(1-x_1)^2]
 \en
and
 \be
 &&\frac{i}{8 P^+ P^{\prime +}}{\rm Tr}\Big[(\not\!\bar P+M_0)\sigma^{i+}(\not\!\bar P'+M'_0)
  \Big(\not\!p_{23}-\frac{\bar P'\cdot p_{23}}{M'_0}\Big)(\not\!  p'_1-m'_1)\gamma^+
  (\not\! p_1-m_1)\Big(\not\!p_{23}-\frac{\bar P\cdot p_{23}}{M_0}\Big)\Big]
 \non\\
 &&\quad=p^2_{23}[(m_1+x_1 M_0) k_{1\bot}^{\prime i}-(m'_1+x_1 M'_0) k_{1\bot}^i]
 \non\\
 &&\quad +(p'_1\cdot \bar P-m'_1 M_0)
               ~[x_{23} k_{23\bot}^i(m_1 + x_1 M'_0)+ x^2_{23} (m_1 q_\bot^i-M'_0 k_{1\bot}^i)
                 +x_1(k_{23\bot}^i+x_{23} q_\bot^i) e_{23}
 \non\\
 &&\qquad       -(x_1 k_{23\bot}^i-x_{23} k_{1\bot}^i) e'_{23}]
 \non\\
 &&\quad     +(p'_1\cdot \bar P'-m'_1 M'_0)
               ~[-x_{23} k_{23\bot}^i(m_1+x_1 M_0)+x^2_{23} M_0 k_{1\bot}^i
                 +x_{23} k_{1\bot}^i e_{23}]
 \non\\
 &&\quad     +(p'_1\cdot \bar P'+m'_1 M'_0)
               ~2[x_{23} k_{23\bot}^i(m_1+x_1 M_0)-x^2_{23} M_0 k_{1\bot}^i]
 \non\\
 &&\quad     +(p_1\cdot \bar P+m_1 M_0)
               ~[-x_{23}(k_{23\bot}^i+ x_{23}q_\bot^i)(m'_1 + x_1 M'_0)+x^2_{23} M'_0 k^{\prime i}_1
                 -x_{23} k_{1\bot}^{\prime i} e'_{23}]
 \non\\
 &&\quad     +(p_1\cdot \bar P'+ m_1 M'_0)
               ~[-x_{23} k_{23\bot}^i(m'_1+x_1 M_0)-x^2_{23} M_0 (x_1 q_\bot^i-k_{1\bot}^{\prime i})
 \non\\
 &&\qquad
               +(x_1 k_{23\bot}^i+x_1 x_{23}q_\bot^i-x_{23} k_{1\bot}^{\prime i})e_{23}-x_1 k_{23\bot}^i e'_{23}]
 \non\\
 &&\quad     +(p_1\cdot p'_1+m_1 m'_1)
               ~[x_{23} k_{23\bot}^i(M_0- M'_0)+x^2_{23} M_0 q_\bot^i
               -(k_{23\bot}^i+x_{23} q_\bot^i) e_{23}+k^i_{23\bot} e'_{23}]
 \non\\
 &&\quad     +(\bar P\cdot \bar P'-M_0 M'_0)
               ~[x_1 x_{23} k_{23\bot}^i (m'_1-m_1)+x^2_{23}(m_1 k_{1\bot}^{\prime i}-m'_1 k_{1\bot}^i-x_1 m_1 q_\bot^i)
 \non\\
 &&\qquad
                -x_1(x_1 k_{23\bot}^i+x_1 x_{23} q_\bot^i-x_{23} k_{1\bot}^{\prime i})e_{23}+
                 x_1(x_1 k_{23\bot}^i-x_{23} k_{1\bot}^i)e'_{23}]
 \non\\
 &&\quad+k_{1\bot}^i\{x_{23}e_{23}[m'_1 M_0+M'_0(2m'_1+x_1 M_0)]
 \non\\
 &&\qquad +x_{23}e'_{23}[M_0^{\prime2}+(m'_1+M_0-x_1 M_0)M'_0+2m'_1M_0]
          -(1-x_1) M'_0 e_{23}^{\prime2}\}
 \non\\
 &&\quad +k_{1\bot}^{\prime i}\{[-x_{23}e_{23}[m_1+M_0+(1-x_1) M'_0]  M_0
          -x_{23}e'_{23}(m_1+x_1 M_0)  M'_0 +(1-x_1) M_0 e_{23}^2\}
  \non\\
  &&\quad +x_{23} q_{\bot}^i\{e_{23}[m'_1(2 m_1+M_0+x_1
  M_0)+x_1(1-x_1) M_0 M'_0]
          -e'_{23} (1-x_1)(m_1+x_1 M_0) M'_0\}
  \non\\
 &&\quad+2k_{23\bot}^i\{e_{23}[m'_1(m_1+M_0)+x_1(1-x_1) M_0 M'_0]
                        -e'_{23} (m_1+x_1 M_0)[m'_1+(1-x_1)
                        M'_0]\}
 \non\\
 &&\quad+e_{23}~e'_{23}
              [(m_1+M_0) k_{1\bot}^{\prime i}-(m'_1+M'_0) k_{1\bot}^i],
 \label{eq:trace2}
 \en
for $i=1$ or 2, where we have used $p_{23}=p_2-p_3=p'_2-p'_3$,
$k_{23\bot}=k_{2\bot}-k_{3\bot}$, $x_{23}=x_2-x_3$,
$P^{\prime+}=P^+$, $x'_1=x_1$ and $\bar P^{(\prime)}\cdot
p_{23}=p^{(\prime)}_1\cdot p_{23}=M^{(\prime)}_0
e^{(\prime)}_{23}$ followed from $m^2_2=m^2_3$ with the
constituents $2,3$ being the spectator light diquarks. By
replacing $m'_1\to -m'_1$, $M'_0\to -M'_0$ in the above traces,
one can obtain the corresponding traces of $g_{1,2}(q^2)$ in a
similar manner.

Traces in Eq.~(\ref{eq:trace2}) can be expressed in terms of the
internal variables $x_i$, $k^{(\prime)}_{i\bot}$, $M'_0$. In doing
so, we need identities beyond Eq.~(\ref{eq:internal1}). It is
useful to note that $\bar P-\bar P'=p_1-p'_1=\bar q$, where $\bar
q^{+}=q^+=0$, $\bar q_\bot= q_\bot$, and $\bar q^2=q^2=-q_\bot^2$,
$\bar q\cdot(x_1 \bar P^{(\prime)}-p_1^{(\prime)})=q_\bot\cdot
k^{(\prime)}_{1\bot}$. We then obtain
 \be
 &&p_1\cdot p'_1=\frac{1}{2}(m_1^2+m_1^{\prime 2}-q^2),\quad
 \bar P\cdot \bar P'_1=\frac{1}{2}(M_0^2+M_0^{\prime 2}-q^2),
 \non\\
 &&p_1\cdot \bar P'=\frac{1}{x_1}[(x_1 \bar
 P^{\prime}-p_1^{\prime})\cdot (p'_1+\bar q)+p'_1\cdot p_1]
 =\frac{m_1^2+x^2_1 M_0^{\prime 2}+(k'_{1\bot}+q_\bot)^2}{2 x_1},
 \non\\
 &&p'_1\cdot \bar P=\frac{1}{x_1}[(x_1 \bar
 P-p_1)\cdot (p_1-\bar q)+p'_1\cdot p_1]
  =\frac{m_1^{\prime 2}+x^2_1 M_0^2+(k_{1\bot}-q_\bot)^2}{2 x_1},
 \non\\
 &&p_{23}^2=k^2_{23}=(k^+_2-k_3^+)(k^-_{2}-k^-_{3})-k_{23\bot}^2,\quad
p^{(\prime)}_1\cdot p_{23}=\bar P^{(\prime)}\cdot
p^{(\prime)}_{23}
 =M_0^{(\prime)}e^{(\prime)}_{23},
  \label{eq:internal2}
 \en
where $p'_{2,3}=p_{2,3}$ and $m_2=m_3$ have been used in the last
identity.

Putting all the pieces together, we obtain the explicit
expressions of the $\P_Q(\6bar)\to \P_{Q'}(\6bar)$ transition form
factors:
 \be
 f_1(q^2)&=&\int \frac{dx_2 d^2 k_{2\bot}}{2 (2\pi)^3}\frac{dx_3 d^2 k_{3\bot}}{2 (2\pi)^3}~
 \frac{\phi^{\prime*}_{00}(\{x'\},\{k'_{\bot}\})~\phi_{00}(\{x\},\{k_{\bot}\})}
 {6\beta_{23}\beta'_{23}\sqrt{[(m_1+x_1 M_0)^2+k_{1\bot}^2][(m'_1+x_1 M'_0)^2+k_{1\bot}^{\prime 2}]}}
 \non\\
&\times&
 \Bigg\{k^2_{23}
       ~[-k_{1\bot}\cdot k'_{1\bot}-(m_1+x_1 M_0)(m'_1+x_1 M'_0)]
 \non\\
 &&     +\frac{x^2_{23}}{4 x^2_1}\{[(m_1+x_1 M_0)^2+k_{1\bot}^2]
                                        [(m'_1+x_1 M'_0)^2+k_{1\bot}^{\prime 2}]
  \non\\
 &&     +x^2_1 [(m_1-m'_1)^2-q^2][(M_0-M'_0)^2-q^2]
 \non\\
 &&     -[(m_1+x_1 M'_0)^2+(k'_{1\bot}+q_\bot)^2][(m'_1+x_1 M_0)^2+(k_{1\bot}-q_\bot)^2]\}
 \non\\
 &&     +e_{23}
        \Big\{-(1-x_1)M_0 e_{23}(m'_1+x_1 M'_0)
 \non\\
 &&     +\frac{x_{23}}{2 x_1}\{[(m_1+x_1 M'_0)^2+(k'_{1\bot}+q_\bot)^2](m'_1+x_1 M_0)
 \non\\
 &&     +[(m'_1+x_1 M_0)^2+(k_{1\bot}-q_\bot)^2](M_0-x_1 M'_0)
 \non\\
 &&     -[(m'_1+x_1 M'_0)^2+k_{1\bot}^{\prime2}](m_1+M_0)
 \non\\
 &&     +x_1[(m_1-m'_1)^2-q^2](M'_0-M_0)
             -x^2_1[(M_0-M'_0)^2-q^2](m'_1+M_0)\}\Big\}
 \non\\
 &&     +e'_{23}
         \Big\{-(1-x_1)M'_0 e'_{23} (m_1+x_1 M_0)
 \non\\
 &&     +\frac{x_{23}}{2 x_1}\{[(m'_1+x_1 M_0)^2+(k_{1\bot}-q_\bot)^2](m_1+x_1 M'_0)
 \non\\
 &&     +[(m_1+x_1 M'_0)^2+(k'_{1\bot}+q_\bot)^2](M'_0-x_1 M_0)
 \non\\
 &&     -[(m_1+x_1 M_0)^2+k_{1\bot}^2](m'_1+M'_0)
 \non\\
 &&     +x_1[(m_1-m'_1)^2-q^2](M_0-M'_0)
             -x^2_1[(M'_0-M_0)^2-q^2](m_1+M'_0)\}\Big\}
 \non\\
  &&+e_{23} e'_{23}~ [k_{1\bot}\cdot k'_{1\bot}+(m_1+M_0)(m'_1+M'_0)-M_0 M'_0(1-x_1)^2]
             \Bigg\},
  \non\\
\frac{f_2(q^2)}{M+M'}&=&-\frac{1}{q^i_\bot}\int \frac{dx_2 d^2
k_{2\bot}}{2 (2\pi)^3}\frac{dx_3 d^2 k_{3\bot}}{2 (2\pi)^3}~
 \frac{\phi^{\prime*}_{00}(\{x'\},\{k'_{\bot}\})~\phi_{00}(\{x\},\{k_{\bot}\})}
 {6\beta_{23}\beta'_{23}\sqrt{[(m_1+x_1 M_0)^2+k_{1\bot}^2][(m'_1+x_1 M'_0)^2+k_{1\bot}^{\prime 2}]}}
 \non\\
&\times&
 \Bigg\{k^2_{23}[(m_1+x_1 M_0) k_{1\bot}^{\prime i}-(m'_1+x_1 M'_0) k_{1\bot}^i]
 \non\\
 &&\quad +\frac{1}{2 x_1}\{[(m'_1-x_1 M_0)^2+(k'_{1\bot}+q_\bot)^2]
               ~[x_{23} k_{23\bot}^i(m_1 + x_1 M'_0)+ x^2_{23} (m_1 q_\bot^i-M'_0 k_{1\bot}^i)
 \non\\
 &&\qquad       +x_1(k_{23\bot}^i+x_{23} q_\bot^i) e_{23}
                -(x_1 k_{23\bot}^i-x_{23} k_{1\bot}^i) e'_{23}]
 \non\\
 &&\quad     +[(m'_1-x_1 M'_0)^2+k_{1\bot}^{\prime2}]
               ~[-x_{23} k_{23\bot}^i(m_1+x_1 M_0)+x^2_{23} M_0 k_{1\bot}^i
                 +x_{23} k_{1\bot}^i e_{23}]
 \non\\
 &&\quad     +[(m'_1+x_1 M'_0)^2+k_{1\bot}^{\prime2}]
               ~2[x_{23} k_{23\bot}^i(m_1+x_1 M_0)-x^2_{23} M_0 k_{1\bot}^i]
 \non\\
 &&\quad     +[(m_1+x_1 M_0)^2+k_{1\bot}^2]
               ~[-x_{23}(k_{23\bot}^i+ x_{23}q_\bot^i)(m'_1 + x_1 M'_0)+x^2_{23} M'_0 k^{\prime i}_1
                 -x_{23} k_{1\bot}^{\prime i} e'_{23}]
 \non\\
 &&\quad     +[(m_1+x_1 M'_0)^2+(k'_{1\bot}+q_\bot)^2]
               ~[-x_{23} k_{23\bot}^i(m'_1+x_1 M_0)-x^2_{23} M_0 (x_1 q_\bot^i-k_{1\bot}^{\prime i})
 \non\\
 &&\qquad
               +(x_1 k_{23\bot}^i+x_1 x_{23}q_\bot^i-x_{23} k_{1\bot}^{\prime i})e_{23}-x_1 k_{23\bot}^i e'_{23}]
 \non\\
 &&\quad     +x_1 [(m_1+m'_1)^2-q^2]
               ~[x_{23} k_{23\bot}^i(M_0- M'_0)+x^2_{23} M_0 q_\bot^i
               -(k_{23\bot}^i+x_{23} q_\bot^i) e_{23}+k^i_{23\bot} e'_{23}]
 \non\\
 &&\quad     +x_1 [(M_0-M'_0)^2-q^2]
               ~[x_1 x_{23} k_{23\bot}^i (m'_1-m_1)+x^2_{23}(m_1 k_{1\bot}^{\prime i}-m'_1 k_{1\bot}^i-x_1 m_1 q_\bot^i)
 \non\\
 &&\qquad
                -x_1(x_1 k_{23\bot}^i+x_1 x_{23} q_\bot^i-x_{23} k_{1\bot}^{\prime i})e_{23}+
                 x_1(x_1 k_{23\bot}^i-x_{23} k_{1\bot}^i)e'_{23}]\}
 \non\\
 &&\quad+k_{1\bot}^i\{x_{23}e_{23}[m'_1 M_0+M'_0(2m'_1+x_1 M_0)]
 \non\\
 &&\qquad +x_{23}e'_{23}[M_0^{\prime2}+(m'_1+M_0-x_1 M_0)M'_0+2m'_1M_0]
          -(1-x_1) M'_0 e_{23}^{\prime2}\}
 \non\\
 &&\quad +k_{1\bot}^{\prime i}\{[-x_{23}e_{23}[m_1+M_0+(1-x_1) M'_0]  M_0
          -x_{23}e'_{23}(m_1+x_1 M_0)  M'_0 +(1-x_1) M_0 e_{23}^2\}
  \non\\
  &&\quad +x_{23} q_{\bot}^i\{e_{23}[m'_1(2 m_1+M_0+x_1
  M_0)+x_1(1-x_1) M_0 M'_0]
          -e'_{23} (1-x_1)(m_1+x_1 M_0) M'_0\}
  \non\\
 &&\quad+2k_{23\bot}^i\{e_{23}[m'_1(m_1+M_0)+x_1(1-x_1) M_0 M'_0]
                        -e'_{23} (m_1+x_1 M_0)[m'_1+(1-x_1)
                        M'_0]\}
 \non\\
 &&\quad+e_{23}~e'_{23}
              [(m_1+M_0) k_{1\bot}^{\prime i}-(m'_1+M'_0)
              k_{1\bot}^i]\Bigg\},
\non\\
  g_{1}(q^2)&=&\int \frac{dx_2 d^2 k_{2\bot}}{2 (2\pi)^3}\frac{dx_3 d^2 k_{3\bot}}{2 (2\pi)^3}~
 \frac{\phi^{\prime*}_{00}(\{x'\},\{k'_{\bot}\})~\phi_{00}(\{x\},\{k_{\bot}\})}
 {6\beta_{23}\beta'_{23}\sqrt{[(m_1+x_1 M_0)^2+k_{1\bot}^2][(m'_1+x_1 M'_0)^2+k_{1\bot}^{\prime 2}]}}
 \non\\
&\times&
 \Big\{{\rm
 the~integrand~of}~f_1(q^2)\Big\}\Big|_{m'_1\to-m'_1,\,M'_0\to-M'_0,\,e'_{23}\to-e'_{23}}~,
 \non\\
 \frac{g_{2}(q^2)}{M+M'}&=&\frac{1}{q^i_\bot}\int \frac{dx_2 d^2 k_{2\bot}}{2 (2\pi)^3}\frac{dx_3 d^2 k_{3\bot}}{2 (2\pi)^3}~
 \frac{\phi^{\prime*}_{00}(\{x'\},\{k'_{\bot}\})~\phi_{00}(\{x\},\{k_{\bot}\})}
 {6\beta_{23}\beta'_{23}\sqrt{[(m_1+x_1 M_0)^2+k_{1\bot}^2][(m'_1+x_1 M'_0)^2+k_{1\bot}^{\prime 2}]}}
 \non\\
&\times&
 \Big\{{\rm
 the~integrand~of}~\frac{f_2(q^2)}{M+M'}\Big\}\Big|_{m'_1\to-m'_1,\,M'_0\to-M'_0,\,e'_{23}\to-e'_{23}}~,
 \label{eq:ff6to6}
 \en
where $i=1$ or 2. It is useful to recall that we have $0\leq
x'_{1,2,3}(=x_{1,2,3})\leq 1$, $x_{23}=x_2-x_3$,
$e^{(\prime)}_{23}=e^{(\prime)}_2-e^{(\prime)}_3$,
$k_{23}=k_2-k_3$, $k'_{1\bot}=k_{1\bot}-(1-x_1)~q_\bot$ and
$k'_{2,3\bot}=k_{2,3\bot}+x_{2,3}~q_\bot$.

\section{Form factors in the Heavy quark limit}

In the heavy quark limit, heavy quark symmetry (HQS) \cite{IW89}
provides model-independent constraints on the form factors. More
precisely, all the heavy-to-heavy baryonic decay form factors are
reduced to some universal Isgur-Wise functions. Therefore, it is
important to study the heavy quark limit behavior of these
physical quantities to check the consistency of calculations.

In the limit of HQS, the weak transitions between antitriplet and
antitriplet or sextet and sextet heavy baryons (i.e. charmed or
bottom baryons) have been studied within the framework of HQET.
For heavy baryons, the two light quarks form either a flavor
symmetric sextet and spin 1 state, or a flavor antisymmetric
antitriplet and spin 0 configuration. As noted in passing, the
four light quarks in the Jaffe-Wilczek model form a flavor
antisextet and $L=1$, $S=0$ state, or a flavor triplet and $L=0$,
$S=0$ state. In the infinite quark mass limit, the heavy quark
spin $S_Q$ decouples from the other degrees of freedom of the
hadron, so that $S_Q$ and the total angular momentum $j$ of the
light quarks (the so-called ``brown muck") are separately good
quantum numbers. Whether the total spin of the light degrees of
freedom comes from the quark spin plus orbital angular momentum or
just from the quark spin is irrelevant. Therefore in the heavy
quark limit, the $\6bar_f$ (${\bf 3}_f$) pentaquark is the analog
of the ${\bf 6}_f$ ($\3bar_f$) heavy baryon. Consequently, the
previous studies of heavy-to-heavy baryon transitions in nineties
using HQET can be generalized to the heavy-to-heavy pentaquark
transitions.

In the heavy quark limit, the ${\bf 3}_f-{\bf 3}_f$
($\6bar_f-\6bar_f$) pentaquark transition is similar to that of
the $\3bar_f-\3bar_f$ (${\bf 6}_f-{\bf 6}_f$) ordinary baryon
transition. Hence, we can write \cite{IW91,Georgi,Yan92}
 \be
 \la \P_{Q'}({\bf 3};v',S'_z)|\bar Q_v\gamma_\mu(1-\gamma_5)
 Q'_{v'}|\P_Q({\bf 3};v,S_z)\ra &=& -\zeta(\omega)\bar
 u(v',S'_z)\gamma_\mu(1+\gamma_5) u(v,S_z)
 \label{eq:HQ3->3}
 \en
and~\cite{Leibovich}
 \be
 \la \P_{Q'}(\6bar;v',S'_z)|\bar Q_v\Gamma
 Q'_{v'}|\P_Q(\6bar;v,S_z)\ra &=& {1\over 3}\left[g^{\mu\nu}\xi_1(\omega)
 -v^\mu v'^\nu\xi_2(\omega)\right]\bar
 u(v',S'_z)\gamma_5(\gamma_\mu+v'_\mu)  \non \\
 &\times& (C^{-1}\Gamma C)^T (\gamma_\nu+v_\nu)\gamma_5\, u(v,S_z),
 \label{eq:HQ6->6compact}
 \en
or explicitly,
 \be    \label{eq:HQ6->6}
 \la \P_{Q'}(\6bar;v',S'_z)|\bar Q_v\gamma_\mu
 Q'_{v'}|\P_Q(\6bar;v,S_z)\ra &=& {1\over 3}\bar
 u(v',S'_z)\Big\{[\omega\gamma_\mu-2(v+v')_\mu]\xi_1(\omega)  \non \\
 &+& [(1-\omega^2)\gamma_\mu-2(1-\omega)(v+v')_\mu]\xi_2(\omega)\Big\}
  u(v,S_z),  \non  \\
  \la \P_{Q'}(\6bar;v',S'_z)|\bar Q_v\gamma_\mu\gamma_5
 Q'_{v'}|\P_Q(\6bar;v,S_z)\ra &=& -{1\over 3}\bar
 u(v',S'_z)\Big\{[\omega\gamma_\mu-2(v-v')_\mu]\xi_1(\omega)  \\
 &+&
 [(1-\omega^2)\gamma_\mu+2(1+\omega)(v-v')_\mu]\xi_2(\omega)\Big\}\gamma_5
  u(v,S_z), \non
 \en
where $h_{Q^{(')}}$ are the dimensionless heavy quark fields,
$\omega\equiv v\cdot v'$, $\zeta,\xi_1$ and $\xi_2$ are three
universal baryonic Isgur-Wise functions with the normalization
$\zeta(1)=\xi_1(1)=1$, and $C$ is the charge conjugation matrix
with $[C^{-1}\gamma_\mu (1-\gamma_5) C]^T=-\gamma_\mu
(1+\gamma_5)$. The normalization of $\xi_2(\omega)$ at zero recoil
is unknown, but the large-$N_c$ approach \cite{Chow} and the quark
model \cite{Chengbottom} predict that $\xi_2(1)=1/2$. From Eqs.
(\ref{eq:HQ3->3}) and (\ref{eq:HQ6->6}) we obtain
 \be \label{eq:FF3->3}
 f_1=g_1=\zeta(\omega), \qquad\qquad f_2=f_3=g_2=g_3=0
 \en
for $\P_{Q}({\bf 3})\to \P_{Q'}({\bf 3})$ transitions and
 \be \label{eq:FF6->6}
 f_1 = F_1+{1\over 2}(M+M')\left({F_2\over M}+{F_3\over
 M'}\right),&& \qquad  g_1 = G_1-{1\over 2}(M-M')\left({G_2\over M}+{G_3\over
 M'}\right), \non \\
 f_2={1\over 2}(M+M')\left({F_2\over M}+{F_3\over
 M'}\right),&& \qquad g_2={1\over 2}(M+M')\left({G_2\over M}+{G_3\over
 M'}\right), \non \\
 f_3={1\over 2}(M+M')\left({F_2\over M}-{F_3\over
 M'}\right), && \qquad g_3={1\over 2}(M+M')\left({G_2\over M}-{G_3\over
 M'}\right),
 \en
with
 \be
 && F_1=G_1=-{1\over 3}[\omega\xi_1+(1-\omega^2)\xi_2], \non \\
 && F_2=F_3={2\over 3}[\omega\xi_1+(1-\omega)\xi_2], \non \\
 && G_2=-G_3={2\over 3}[\omega\xi_1-(1+\omega)\xi_2]
 \en
for $\P_{Q}(\6bar)\to \P_{Q'}(\6bar)$ transitions. It is
straightforward to show that at zero recoil $q^2=q_m^2\equiv
(M-M')^2$ \cite{Chengbottom} \be \label{eq:3to3zerorecoil}
 f_1(q_m^2)=g_1(q_m^2)=1, \quad f_{2,3}(q^2_m)=g_{2,3}(q^2_m)=0
 \en
for $\P_{Q}({\bf 3})\to \P_{Q'}({\bf 3})$, and
 \be \label{eq:6to6zerorecoil}
 && f_1(q_m^2)=-{1\over 3}\left[1-(M+M')\left({1\over M}+{1\over
 M'}\right)\right], \non \\
 && f_2(q_m^2)={1\over 3}(M+M')\left({1\over M}+{1\over M'}\right),
 \qquad f_3(q_m^2)={1\over 3}(M+M')\left({1\over M}-{1\over
 M'}\right), \non \\
 && g_1(q_m^2)=-{1\over 3}, \qquad\quad g_2(q_m^2)=g_3(q_m^2)=0
 \en
for $\P_{Q}(\6bar)\to \P_{Q'}(\6bar)$. The form factors
(\ref{eq:3to3zerorecoil}) and (\ref{eq:6to6zerorecoil})  are model
independent and should be respected by any model calculations.

We now consider the heavy quark limit of transition form factors
obtained in the previous section and begin with the $\P_{Q}({\bf
3})\to\P_{Q'}({\bf 3})$ transition. Under the HQ limit, we have
 \be
 &&Q|\P_{Q} (P,S_z)\ra \to \sqrt{m_Q}\,
   Q_v|\P_Q (v,S_z)\ra,\quad
 u(\bar P,S_z)\to \sqrt{m_Q}\,
 u(v,S_z),\quad
 \not\!p_1+m_1\to m_Q(\not\!v+1),
 \non\\ &&\qquad\qquad\sqrt{\frac{e_1 e_2 e_3}{x_1 x_2 x_3 M_0}}\to m_Q \sqrt{\frac{v\cdot p_2~v\cdot
 p_3}{X_2 X_3}},\quad
 \phi_{00}\to \frac{m_Q}{\sqrt{X_2 X_3}}\Phi_{00},
\label{eq:HQlimit}
 \en
with $X_{2,3}\equiv m_Q x_{2,3}$ and
 \be
 \Phi_{00}
 &=&16 \left({\pi^2\over{\beta^{2}_1\beta^{2}_{23}}}\right)^{3\over{4}}
 \sqrt{v\cdot p_2~v\cdot
 p_3}~{\rm exp}\Bigg\{-\left(\frac{1}{2\beta_1^2}-\frac{1}{8\beta_{23}^2}\right)
 2\Big(\vec k_2\cdot \vec k_3\Big) \non \\
 && -\left(\frac{1}{2\beta_1^2}+\frac{1}{8\beta_{23}^2}\right)
 \left[(v\cdot p_2)^2-m_2^2+(v\cdot p_3)^2-m_3^2\right]
 \Bigg\}
 \label{eq:HQreduction}
 \en
and similar relations for $u(\bar P',S'_z)$, etc. Since $x_2$ and
$x_3$ are of  order $\Lambda_{\rm QCD}/m_Q$ in the $m_Q\to\infty$
limit, it is clear that $X_2$ and $X_3$ are of order $\Lambda_{\rm
QCD}$. For on-shell $p_i$ $(i=2,3)$ we have $p_i^- = (p_{i\bot}^2
+ m_i^2)/p_i^+$ and
 \begin{equation}
    v \cdot p_i = {1\over 2X_i} \Big( p_{i\bot}^2 + m_i^2
        + X_i^2 \Big) \, .
 \end{equation}
Applying Eqs. (\ref{eq:HQlimit}) and (\ref{eq:3->3}), we obtain
 \be
  &&\la \P_{Q'}({\bf 3};v',S'_z)|\bar Q_v\gamma^\mu(1-\gamma_5)
 Q'_{v'}|\P_Q({\bf 3};v,S_z)\ra   \\
&&=\int \frac{dX_2 d^2 k_{2\bot}}{2 (2\pi)^3 X_2}\frac{dX_3 d^2
  k_{3\bot}}{2 (2\pi)^3 X_3}~
 \Phi^{\prime*}_{00}(zX_2,p_{2\bot}^2;zX_3,p_{3\bot}^2)~\Phi_{00}(X_2,p_{2\bot}^2;X_3,p_{3\bot}^2)
 \non \\
 && ~\times\bar u(v',S'_z) 
 \gamma^\mu(-1-\gamma_5) 
 u(v,S_z), \non
 \en
where use of $z\equiv X'_2/X_2=X'_3/X_3$ has been made. By
comparing the above equation with Eq. (\ref{eq:HQ3->3}), we find
 \be
 \zeta(\omega)=\int \frac{dX_2 d^2 k_{2\bot}}{2 (2\pi)^3 X_2}\frac{dX_3 d^2
  k_{3\bot}}{2 (2\pi)^3 X_3}~
 \Phi^{\prime*}_{00}(zX_2,p_{2\bot}^2;zX_3,p_{3\bot}^2)~\Phi_{00}(X_2,p_{2\bot}^2;X_3,p_{3\bot}^2).
 \label{eq:zeta}
 \en
It can be easily seen that $\zeta(1)=1$ from Eqs.~(\ref{momnor})
and (\ref{eq:HQreduction}). Note that $z$ is related to $\omega$
via \cite{CCH}
\begin{equation}
    z \rightarrow z_{\pm} = \omega \pm \sqrt{\omega^2
        - 1}~,  ~~~~~ z_+ = {1 \over z_-} \, ,
\end{equation}
with the + $(-)$ sign corresponding to $v^3$ greater (less) than
$v'^3$. Note that $v^3$ greater (less) than $v'^3$ corresponds the
daughter meson recoiling in the negative (positive) $z$ direction
in the rest frame of the parent meson.  One can check that
$\zeta(\omega)$ remains the same under the replacement of $z\to
1/z$. This indicates that the Isgur-Wise function thus obtained is
independent of the recoiling direction, namely, it is truly
Lorentz invariant.

As to the $\P_{Q}({\bf 6})\to\P_{Q'}({\bf 6})$ case, the heavy
quark limit of Eq. (\ref{eq:6->6}) leads to
 \be
 &&\la \P_{Q'}(\6bar;v',S'_z)|\bar Q_v\gamma^\mu (1-\gamma_5)
 Q'_{v'}|\P_{Q}(\6bar;v,S_z)\ra
 \non\\
&&\qquad\qquad=-\int \frac{dX_2 d^2 k_{2\bot}}{2 (2\pi)^3
X_2}\frac{dX_3 d^2 k_{3\bot}}{2 (2\pi)^3 X_3}~
 \frac{\Phi^{\prime*}_{00}~\Phi_{00}}
 {24\beta_{23}\beta'_{23}}
 ~(p'_2-p'_3)_\rho(p_2-p_3)_\sigma
 \non\\
 &&\qquad\qquad~\times~ \bar u(v',S'_z) \gamma_5 (\gamma^\rho+v^{\prime\rho})(\not\!
 v'+1)\gamma^\mu(-1-\gamma_5)(\not\!
 v+1) (\gamma^\sigma+v^\sigma) \gamma_5 u(v,S_z)
 \non\\
 &&\qquad\qquad=-\int \frac{dX_2 d^2 k_{2\bot}}{2 (2\pi)^3 X_2}\frac{dX_3 d^2 k_{3\bot}}{2 (2\pi)^3 X_3}~
 \frac{\Phi^{\prime*}_{00}~\Phi_{00}}
 {6\beta_{23}\beta'_{23}}
 ~(p'_2-p'_3)_\rho(p_2-p_3)_\sigma
 \non\\
 &&\qquad\qquad~\times~ \bar u(v',S'_z) \gamma_5 (\gamma^\rho+v^{\prime\rho})\gamma^\mu(-1-\gamma_5)
    (\gamma^\sigma+v^\sigma) \gamma_5 u(v,S_z),
 \label{eq:6->6HQ}
 \en
where uses of Eqs. (\ref{eq:HQlimit}-\ref{eq:HQreduction}), $\bar
u'\gamma_5(\not\! v'+1)(\gamma^\rho+v^{\prime\rho})=2\bar
u'\gamma_5(\gamma^\rho+v^{\prime \rho})$ and $(\not\!
v+1)(\gamma^\sigma+v^\sigma) \gamma_5
u=2(\gamma^\sigma+v^\sigma)\gamma_5 u$ have been made. To proceed
we need to express $(p'_2-p'_3)_\rho(p_2-p_3)_\sigma$ in terms of
$g_{\rho\sigma}$, $v_\rho v'_\sigma$, $v'_\rho v_\sigma$, $v_\rho
v_\sigma$ and $v'_\rho v'_\sigma$ via the general expression
 \be \label{eq:coefficients}
 && \int \frac{dX_2 d^2 k_{2\bot}}{2 (2\pi)^3
X_2}\frac{dX_3 d^2 k_{3\bot}}{2 (2\pi)^3 X_3}~
 \Phi^{\prime*}_{00}~\Phi_{00}~(p'_2-p'_3)_\rho(p_2-p_3)_\sigma
 \non\\
 &=& \int d^3p_2\,d^3p_3~f(v\cdot p_2)f(v\cdot p_3)f(v'\cdot p_2)
  f(v'\cdot p_3) (p'_2-p'_3)_\rho(p_2-p_3)_\sigma \non \\
 &=&  a_1 g_{\rho\sigma}+a_2 v_\rho v'_\sigma+a_3v'_\rho v_\sigma
                 +a_4 v_\rho v_\sigma+a_5 v'_\rho v'_\sigma.
 \en
Applying this to Eq.~(\ref{eq:6->6HQ}), it is easily seen that the
$a_3$, $a_4$ and $a_5$ terms do not make contributions to the
$\P_{Q}(\6bar)\to\P_{Q'}(\6bar)$ transition owing to the relation
$\bar u'\gamma_5(\not\! v'+1)=({\not\! v}+1)\gamma_5 u=0$. As we
shall see shortly, the only relevant coefficients $a_1$ and $a_2$
are related to the Isgur-Wise functions $\xi_1$ and $\xi_2$,
respectively. The coefficients $a_i$ can be solved by contracting
the left and right hand sides of Eq.~(\ref{eq:coefficients}) with
$g^{\rho\sigma},v^\rho v^{\prime\sigma},v'^\rho v^\sigma,v^\rho
v^{\sigma},v'^\rho v'^\sigma$. Noting that $p_{23}\equiv p_2-p_3$
with $p_{2,3}=p'_{2,3}$ in our case, we obtain
 \be
  a_1 &=& \int d^3p_2\,d^3p_3~f(v\cdot p_2)f(v\cdot p_3)f(v'\cdot p_2)
                 f(v'\cdot p_3)\non\\
 &\times& \frac{1}{2(1-\omega^2)}\Big\{(1-\omega^2)p_{23}^2
                            -(v\cdot p_{23})^2+2\omega~v\cdot p_{23}~v'\cdot p_{23}-(v'\cdot
                            p_{23})^2\Big\},
                            \non\\
    a_2 &=& \int d^3p_2\,d^3p_3~f(v\cdot p_2)f(v\cdot p_3)f(v'\cdot p_2)
                 f(v'\cdot p_3)    \\
 &\times& \frac{1}{2(1-\omega^2)^2}\Big\{\omega(1-\omega^2)p^2_{23}+2(1+2 \omega^2) v\cdot p_{23}
 ~v'\cdot p_{23}-3\omega[(v\cdot p_{23})^2+(v'\cdot
 p_{23})^2]\Big\}.
  \non
 \en
Comparing Eq.~(\ref{eq:6->6HQ}) with Eq.~(\ref{eq:HQ6->6compact}),
the HQ limit expression of the $\P_Q(\6bar)\to \P_{Q'}(\6bar)$
transition matrix element, we obtain
 \be
 \xi_1(\omega)&=&-\int \frac{dX_2 d^2 k_{2\bot}}{2 (2\pi)^3
 X_2}\frac{dX_3 d^2k_{3\bot}}{2 (2\pi)^3 X_3}~
 {\Phi^{\prime*}_{00}~\Phi_{00}\over{4\beta_{23}\beta'_{23}}}{1\over{(1-\omega^2)}} \non \\
 &\times& \Big[(1-\omega^2)p_{23}^2
                            -(v\cdot p_{23})^2+2\omega~v\cdot p_{23}~v'\cdot p_{23}-(v'\cdot p_{23})^2\Big]
 \label{eq:xi1}
 \en
and
 \be
 \xi_2(\omega)&=&\int \frac{dX_2 d^2 k_{2\bot}}{2 (2\pi)^3
 X_2}\frac{dX_3 d^2k_{3\bot}}{2 (2\pi)^3 X_3}~
 {\Phi^{\prime*}_{00}~\Phi_{00}\over{4\beta_{23}\beta'_{23}}}{1\over{
 (1-\omega^2)^2}}\non \\
 &\times&\Big\{(\omega-\omega^3)p^2_{23}+2(1+2 \omega^2) v\cdot p_{23}
 ~v'\cdot p_{23}-3\omega[(v\cdot p_{23})^2+(v'\cdot
 p_{23})^2]\Big\}.
 \label{eq:xi2}
 \en
At first sight, it appears that the second IW function $\xi_2$ is
divergent at zero recoil. We shall see below that it is not the
case.

We proceed to calculate the normalization of $\xi_{1,2}(\omega)$
at $\omega=1$. Since $p'_{23\rho} p_{23\sigma}$ in
Eq.~(\ref{eq:6->6HQ}) can be replaced by $\hat p'_{23\rho} \hat
p_{23\sigma}$ with $\hat p^{(\prime)}_{23}\equiv p_{23}-
(v^{(\prime)}\cdot p_{23})~v^{(\prime)}$, the IW functions
$\xi_{1,2}$ can be recast to
 \be
 \xi_1(\omega)&=&-\int \frac{dX_2 d^2 k_{2\bot}}{2 (2\pi)^3
 X_2}\frac{dX_3 d^2k_{3\bot}}{2 (2\pi)^3 X_3}~
 {\Phi^{\prime*}_{00}~\Phi_{00}\over{4\beta_{23}\beta'_{23}}}{1\over{(1-\omega^2)}} 
 \Big[(1-\omega^2)\hat p'_{23}\cdot \hat p_{23}
                            +\omega~v\cdot \hat p'_{23}~v'\cdot \hat
                            p_{23}\Big],\\
 \xi_2(\omega)&=&\int \frac{dX_2 d^2 k_{2\bot}}{2 (2\pi)^3
 X_2}\frac{dX_3 d^2k_{3\bot}}{2 (2\pi)^3 X_3}~
 {\Phi^{\prime*}_{00}~\Phi_{00}\over{4\beta_{23}\beta'_{23}}}{1\over{
 (1-\omega^2)^2}}
 \Big[(\omega-\omega^3)\hat p'_{23}\cdot\hat p_{23}+(2+\omega^2) v\cdot \hat p'_{23}
 ~v'\cdot \hat p_{23}\Big].
 \non
 \en
Since $\hat p^{(\prime)}_{23}=(0,\vec k^{(\prime)}_{23})$ in the
$\P_{Q^{(')}}$ rest frame and $v^{(\prime)}\cdot \hat
p^{(\prime)}_{23}=0$, it follows that $v\cdot \hat p'_{23}
v'\cdot\hat p_{23}=(1-\omega^2) \vec k'_{23}\cdot\hat v~\vec
k_{23}\cdot\hat v$, where $\hat v=\vec v/|\vec v|$ with $\vec v$
in the $\P_{Q'}$ rest frame.\footnote{Since $v'=(1,\vec 0)$,
$v=(w,\hat v\sqrt{w^2-1})$ and $\hat p^\prime_{23}=(0,\vec
k^\prime_{23})$ in the $\P_{Q'}$ rest frame, the corresponding
quantities in the $\P_Q$ rest frame are $v=(1,\vec 0)$,
$v'=(w,-\hat v\sqrt{w^2-1})$ and $\hat p_{23}=(0,\vec k_{23})$.}
We then have
 \be
 \xi_1(\omega)&=&-\int \frac{dX_2 d^2 k_{2\bot}}{2 (2\pi)^3
 X_2}\frac{dX_3 d^2k_{3\bot}}{2 (2\pi)^3 X_3}~
 {\Phi^{\prime*}_{00}~\Phi_{00}\over{4\beta_{23}\beta'_{23}}} 
 \Big(\hat p'_{23}\cdot \hat p_{23}
                            +\omega~\vec
k'_{23}\cdot\hat v~\vec k_{23}\cdot\hat v\Big),
 \en
with $\vec k^{(')}_{23}=\vec k^{(')}_2-\vec k^{(')}_3$, $\vec
k'_{2,3\bot}=\vec k_{2,3\bot}+x_{2,3}\,q_\bot=\vec k_{2,3\bot}$
\footnote{In the infinite quark mass limit, the IW functions can
be evaluated directly in the time-like region by choosing the
frame where $q_\bot=0$ \cite{CCH}.} and
 \be
 k_{2,3z}={X_{2,3}\over 2}-{m_{2,3}^2+k^2_{2,3\bot}\over
 2X_{2,3}}.
 \en

For a further simplification of $\xi_1$, we notice that
$p'_{23}=(0,\vec k'_{23})$, $p_{23}=(\sqrt{\omega^2-1}~\hat v\cdot
\vec k_{23},~\vec k_{23}+(\omega-1)\hat v~\hat v\cdot k_{23})$ in
the $\P_{Q'}$ rest frame. The above equation for $\xi_1$ is then
reduced to
 \be
 \xi_1(\omega)&=&\int \frac{dX_2 d^2 k_{2\bot}}{2 (2\pi)^3
 X_2}\frac{dX_3 d^2k_{3\bot}}{2 (2\pi)^3 X_3}~
 {\Phi^{\prime*}_{00}~\Phi_{00}\over{\beta_{23}\beta'_{23}}} 
 \Big[\frac{\vec k'_{23}}{2}\cdot \frac{\vec k_{23}}{2}-\Big(\frac{\vec k'_{23}}{2}\cdot\hat
 v\Big)\Big( \frac{\vec k_{23}}{2}\cdot\hat
 v\Big)\Big].
  \en
At zero recoil,
 \be
 \xi_1(1)=\int \frac{dX_2 d^2 k_{2\bot}}{2 (2\pi)^3
 X_2}\frac{dX_3 d^2k_{3\bot}}{2 (2\pi)^3 X_3}~
 {\Phi^{\prime*}_{00}~\Phi_{00}\over{\beta_{23}\beta'_{23}}} 
 {(\vec k_{23})^2\over 6}=1,
 \en
where we have made use of the rotational invariance of $\vec
k'_{23}\cdot\hat v~\vec k_{23}\cdot\hat v\to \vec k'_{23}\cdot\vec
k_{23}/3$ and  Eq.~(\ref{momnor}). Likewise, for $\xi_2(\omega)$
we have
 \be
&&{1\over{
 (1-\omega^2)^2}}
 \Big[(\omega-\omega^3)\hat p'_{23}\cdot\hat p_{23}+(2+\omega^2) v\cdot \hat p'_{23}
 ~v'\cdot \hat p_{23}\Big],\non\\
 &&={1\over{
 (1-\omega^2)}}
 \Big[\omega\hat p'_{23}\cdot\hat p_{23}+(2+\omega^2) \vec
k'_{23}\cdot\hat v~\vec k_{23}\cdot\hat v\Big].
 \en
Applying the relation $p'_{23}\cdot p_{23}=-\vec k'_{23}\cdot \vec
k_{23}+(\vec k'_{23}\cdot \hat v)(\vec k_{23}\cdot \hat v)$ to the
right hand side of the above equation leads to
 \be
{1\over{
 (1-\omega^2)}}
 \Big[-\omega\vec k_{23}\cdot\vec k'_{23}-\omega(\omega-1)\vec
k'_{23}\cdot\hat v~\vec k_{23}\cdot\hat v+(2+\omega^2) \vec
k'_{23}\cdot\hat v~\vec k_{23}\cdot\hat v\Big],
 \en
which reduces to ${2}\vec k_{23}\cdot\vec k'_{23}/[{3(1+\omega)}]$
in the $\omega\to 1$ limit by using the argument of rotational
invariance. Therefore, we have $\xi_2(1)=1/2$, in agreement with
the the large-$N_c$ approach \cite{Chow} and the quark model
predictions \cite{Chengbottom}.

In short, the heavy-to-heavy pentaquark transition form factors in
the heavy quark limit can be expressed in terms of three universal
Isgur-Wise functions, as shown in Eqs. (\ref{eq:FF3->3}) and
(\ref{eq:FF6->6}). The fact that these IW functions have the
correct normalization at zero recoil indicates the consistency of
our light-front model calculations.

\section{Weak decays of Heavy Pentaquark baryons}

In this section we will perform the numerical calculations of
various pentaquark transition form factors shown in Sec. III and
the Isgur-Wise functions in Sec. IV. We then proceed to estimate
the decay rates of the some weak decays of heavy pentaquarks such
as $\Theta_b^+\to\Theta_c^0\pi^+,\,\Theta_c^0\rho^+$,
$\Theta_c^0\to\Theta^+\pi^-,\,\Theta^+\rho^-$, $\Sigma'^+_{5b}\to
\Sigma'^0_{5c}\pi^+,\,\Sigma'^0_{5c}\rho^+$ and $\Sigma'^0_{5c}\to
N^+_8\pi^+,\,N^+_8\rho^+$ with $N_8^+$ being an odd-parity octet
pentaquark.

The input parameters $m_{[qq']}$, $m_q$, $\beta_1$ (for the heavy
quark) and $\beta_{23}$ (for the diquark pair) [see Eq.
(\ref{eq:phi})] are summarized in Table~\ref{tab:input}. The quark
masses are taken from \cite{CCH,CC}. For our convenience, the
pentaquark masses are fixed to be 6 and 3 GeV for $\P_b$ and
$\P_c$, respectively. For light anti-decuplet and octet pentaquark
masses, we take
$m_{\Theta}=1.54$~GeV~\cite{LEPS,Diana,CLAS1,Saphir,ITEP,CLAS2,Hermes,SVD,COSY,ZEUS}
and $m_{N_8}=1.46$~GeV~\cite{Zhang}. Note that the $\beta$
parameters are of order $\Lambda_{\rm QCD}$ and the diquark
carries the same color charge as the anti-quark. Since the diquark
pair acts like ${\bf 3}_c$, the $\bar Q$--$\{[ud][ud]\}$ system
can be considered as the analog of the heavy meson $\bar Q$--$q$.
Therefore, it is plausible to assume that
$\beta_{1b}:\beta_{1c}:\beta_{1s}\sim\beta_{B}:\beta_{D}:\beta_{K}$.
The $\beta_{23[ud]}$ parameters for the diquark pair are roughly
estimated as $\propto\beta_{\pi,\rho}$ for $\P_Q({\bf 3})$ and
$\propto\beta_{a_1}$ for $\P_Q(\6bar)$. The numerical values of
$\beta_{B,D,K,\pi,\rho,a_1}$ are taken from \cite{CCH,CC}. As we
shall see shortly, by using these input parameters, the obtained
$\Sigma'_{5b}\to\Sigma'_{5c}$ transition form factors $f_1(0)$,
$g_1(0)$ are close to their counterparts (in the sense of
$SU_f(3)$ representation) in  the $\Lambda_b\to\Lambda_c$
transition \cite{Cheng97a}.

\begin{table}[b]
\caption{\label{tab:input} The pentaquark masses and input
parameters $m_{[qq']}$, $m_q$ and $\beta$'s (in units of GeV)
appearing in the Gaussian-type wave function (\ref{eq:wavefn}).}
\begin{ruledtabular}
\begin{tabular}{cccccccccccc}
          $m_{\P_b}$
          &$m_{\P_c}$
          & $m_{[ud]}$
          & $m_{[us]}$
          & $m_s$
          & $m_b$
          & $m_c$
          & $\beta_{1b}$
          & $\beta_{1c}$
          & $\beta_{1s}$
          & $\beta_{23[ud]}^{(s-wave)}$
          & $\beta_{23[ud]}^{(p-wave)}$
          \\
\hline    6
          & 3
          & $0.40$
          & $0.56$
          & $0.45$
          & $4.4$
          & $1.3$
          & $0.65$
          & $0.58$
          & $0.48$
          & $0.38$
          & $0.38$
\end{tabular}
\end{ruledtabular}
\end{table}

\begin{figure}[t!]
\centerline{
            {\epsfxsize3.5 in \epsffile{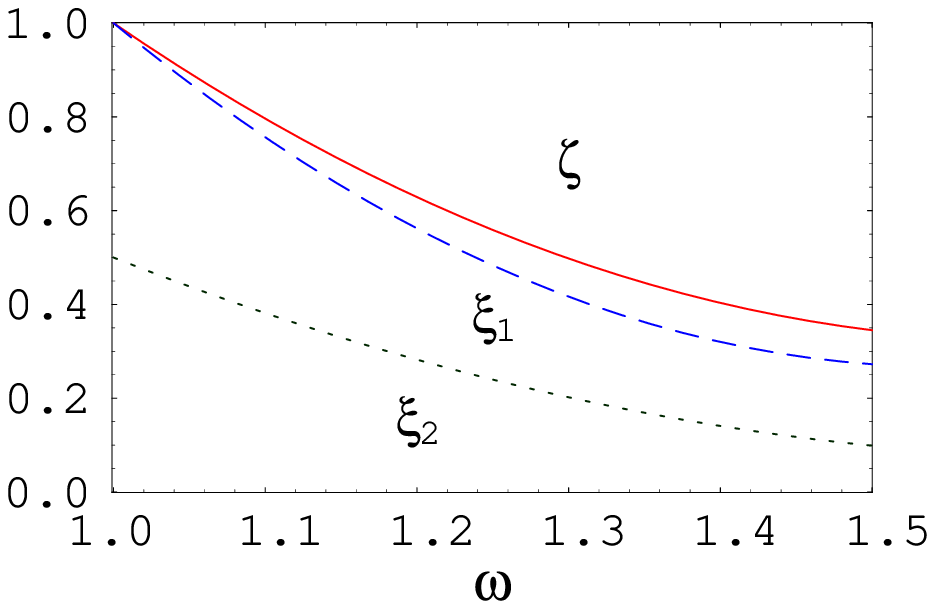}}}
\caption{The Isgur-Wise functions $\zeta$, $\xi_1$ and $\xi_2$ as
a function of $\omega$.} \label{fig:IW} 
\end{figure}

To perform numerical calculations of the IW functions  we choose
$\beta_{1Q}^\infty=0.65$ GeV and $\beta_{23}^\infty=0.38$ GeV as
those in Table \ref{tab:input}. The IW functions (\ref{eq:zeta}),
(\ref{eq:xi1}) and (\ref{eq:xi2}) can be fitted nicely to the form
 \be
 f(\omega)=f(1)\left[1-\rho^2(\omega-1)+{\sigma^2\over 2}(\omega-1)^2\right],
 \en
and it is found that (see Fig. \ref{fig:IW})
 \be
 \zeta(\omega) &=& 1-2.22(\omega-1)+1.82(\omega-1)^2, \non \\
 \xi_1(\omega) &=& 1-2.68(\omega-1)+2.45(\omega-1)^2, \non \\
 \xi_2(\omega) &=&
 0.5\left[1-2.56(\omega-1)+1.91(\omega-1)^2\right].
 \en

We next turn to the form factors. Recall that
$\Sigma'_{5b}\to\Sigma'_{5c}$ is a $\P_{b}({\bf 3})\to\P_{c}({\bf
3})$ transition and $\Theta_{b}\to\Theta_{c}$ is a
$\P_{b}({\6bar})\to\P_{c}({\6bar})$ one. It is straightforward to
obtain their transition form factors from Eqs.~(\ref{eq:ff3->3})
and (\ref{eq:ff6to6}).\footnote{For $f_2(q^2)$ and $g_2(q^2)$, we
take $q^i_\bot=(\sqrt{-q^2},0)$ in Eqs.~(\ref{eq:ff3->3}) and
(\ref{eq:ff6to6}).}
Since the flavor wave functions of $N^+_8(\bar s[ud][us]_-)$ and
$\Theta(\bar s [ud][ud]_+)$ are similar to $\Sigma'_{5c}(\bar s
[ud][ud]_-)$ and $\Theta_c(\bar s [ud][ud]_+)$, respectively, with
$\bar c$ replaced by $\bar s$, the $\Sigma'_{5c}\to N^+_8$ and
$\Theta_{b}\to\Theta_{c}$ transition form factor formula are
similar to Eqs.~(\ref{eq:ff3->3}) and (\ref{eq:ff6to6}),
respectively.
As our calculation of form factors is done in the $q^+=0$ frame
where $q^2\leq 0$, we shall follow~\cite{Jaus96} to analytically
continue the form factors to the timelike region.\footnote{Since
the heavy quark-pair creation is forbidden in the $m_Q\to\infty$
limit, the so-called $Z$-graph which must be incorporated in the
form-factor calculations in order to maintain covariance
\cite{CCHZ} is no longer a problem in the reference frame where
$q^+\geq 0$. This allows us to compute the Isgur-Wise functions
directly in the timelike region.}
To proceed, we find that the momentum dependence of the form
factors in the spacelike region can be well parameterized and
reproduced in the three-parameter form:
 \be \label{eq:FFpara}
 F(q^2)&=&\,{F(0)\over (1-q^2/M^2)
 [1-a(q^2/M^2)+b(q^2/M^2)^2]} \label{eq:FFpara1}
 \en
for $\P_Q\to \P_{Q'}$ transitions. The parameters $a$, $b$ and
$F(0)$ are first determined in the spacelike region. We then
employ this parametrization to determine the physical form factors
at $q^2\geq 0$. The parameters $a,b$ are expected to be of order
${\cal O}(1)$.

\begin{figure}[t!]
\centerline{
            {\epsfxsize3 in \epsffile{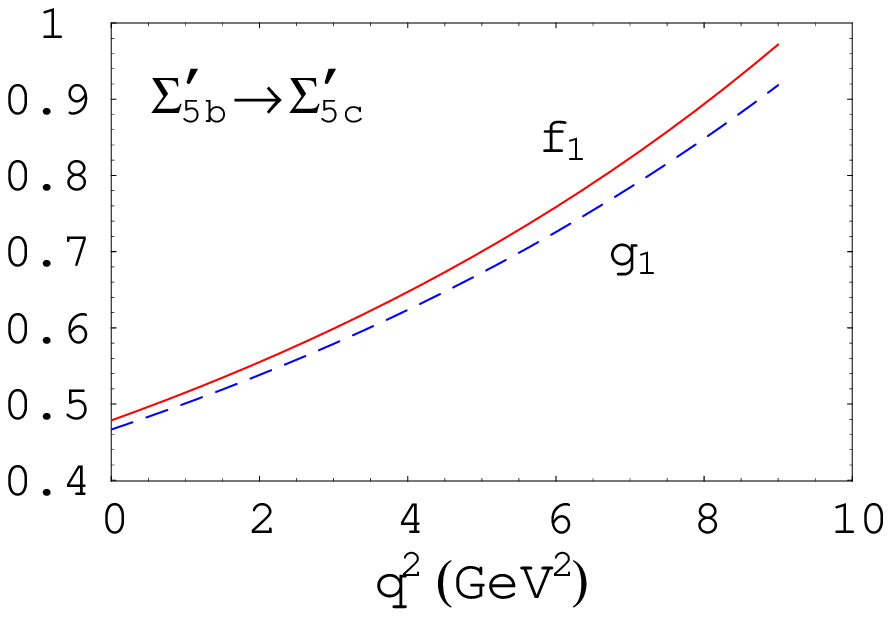}}
            {\epsfxsize3 in \epsffile{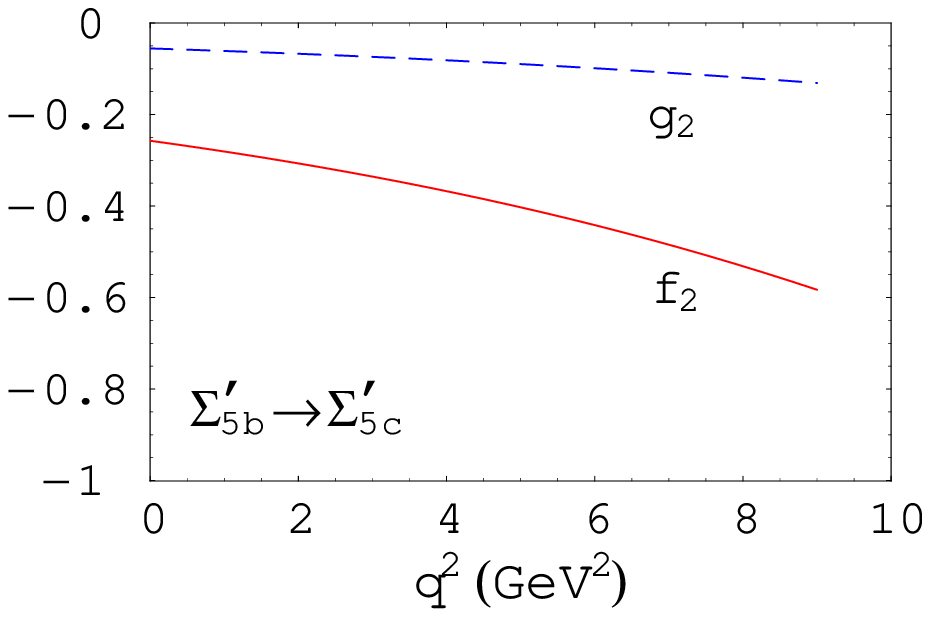}}}
\smallskip
\centerline{
            {\epsfxsize3 in \epsffile{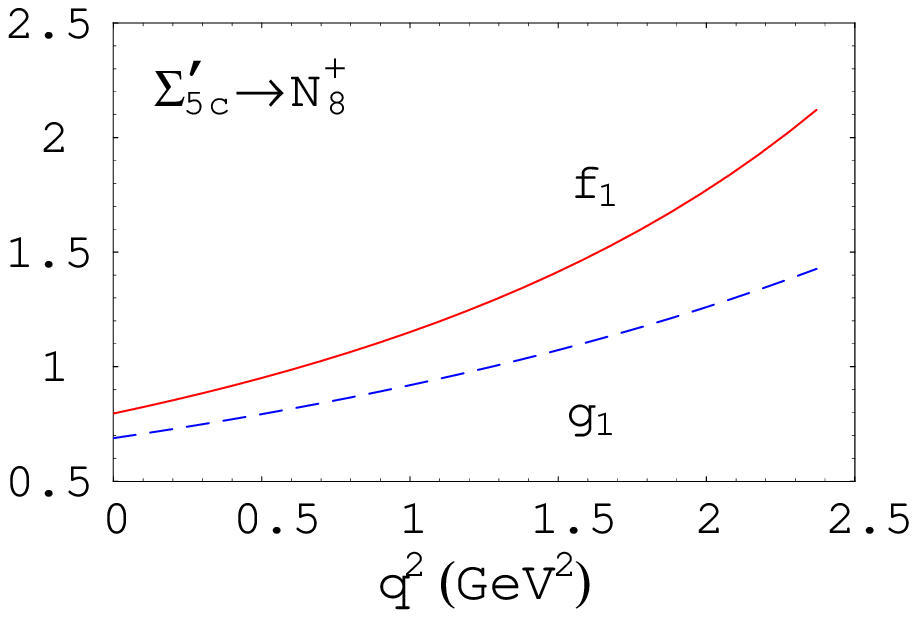}}
            {\epsfxsize3 in \epsffile{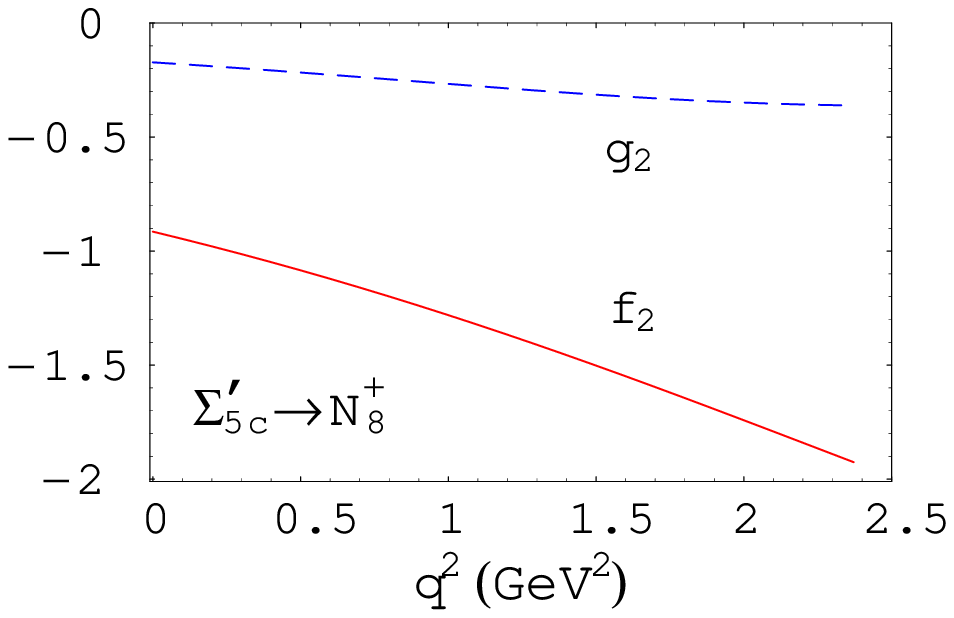}}}
\smallskip
\centerline{
            {\epsfxsize3 in \epsffile{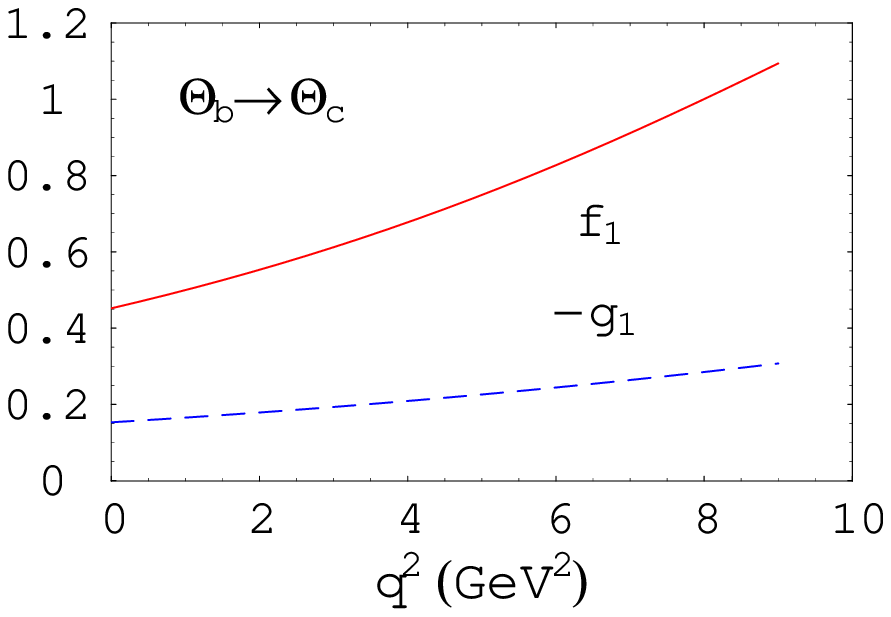}}
            {\epsfxsize3 in \epsffile{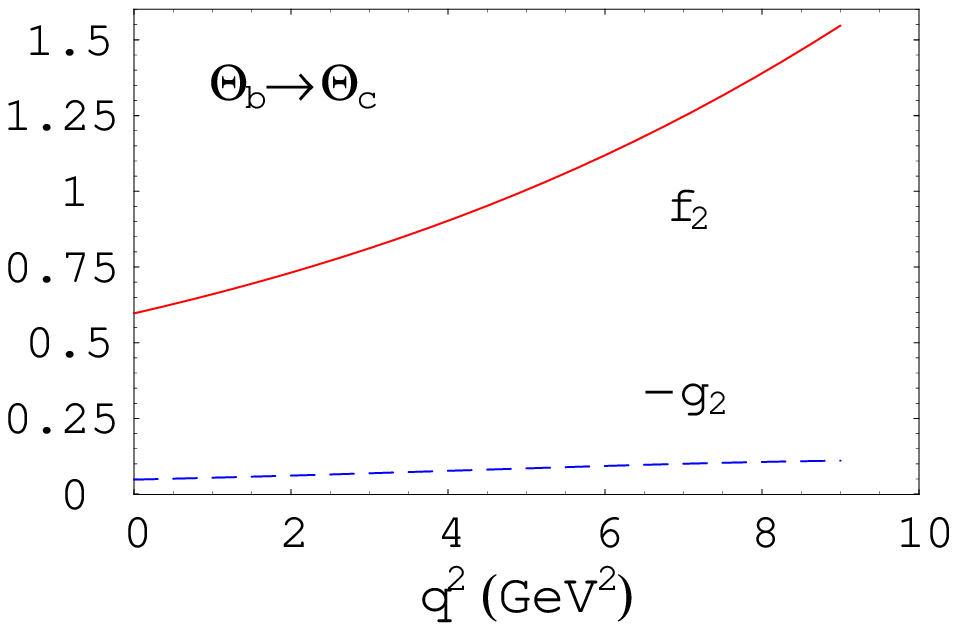}}}
\smallskip
\centerline{
            {\epsfxsize3 in \epsffile{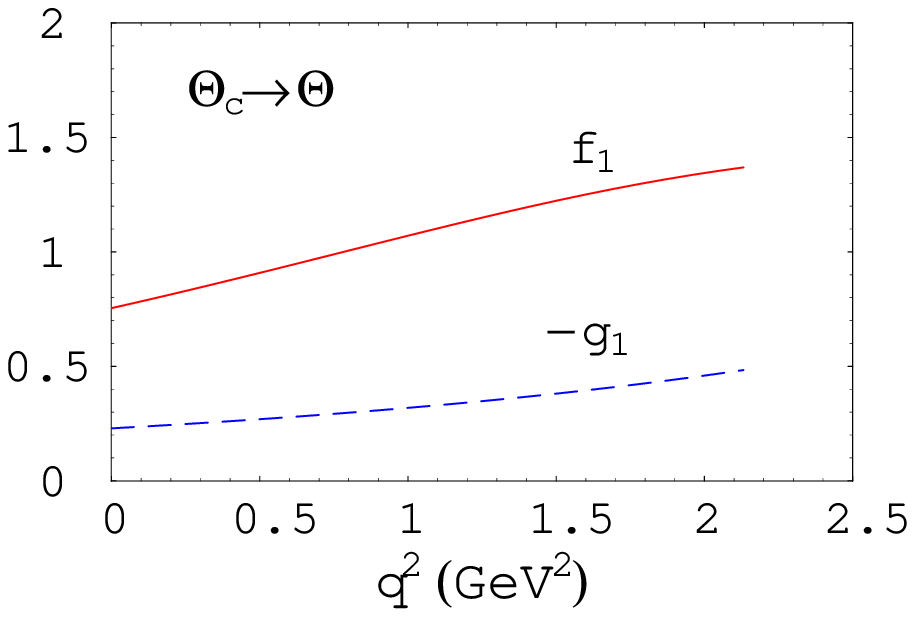}}
            {\epsfxsize3 in \epsffile{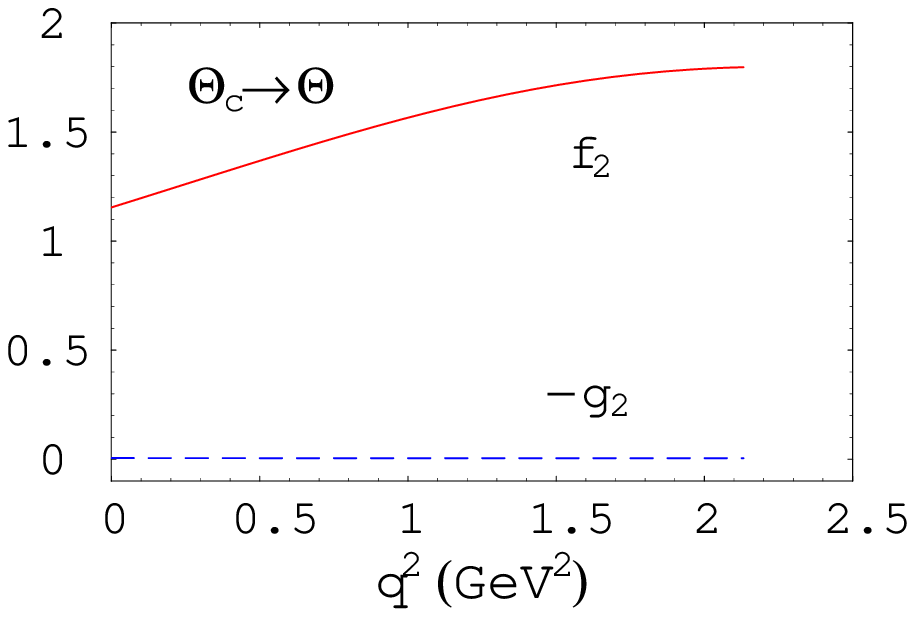}}}
\caption{Form factors $f_{1,2}(q^2)$ and $g_{1,2}(q^2)$ for
$\Sigma^{\prime+}_{5b}\to\Sigma^{\prime0}_{5c}$,
$\Sigma^{\prime0}_{5c}\to N^+_{8}$ and,
$\Theta_{b(c)}\to\Theta_{c(s)}$ transitions.} \label{fig:figi} 
\end{figure}

\begin{table}[t!]
\caption{\label{tab:fg} The transition form factors for various
$\P_Q\to\P_{Q'}$ transitions. Note that $F_{\rm HQ}(q^2_{\rm
max})$ are the form factors at zero recoil in the heavy quark
limit.}
\begin{ruledtabular}
\begin{tabular}{cccccc|cccccc}
           $F^{\P_Q\P_{Q'}}$
          & $F(0)$
          & $F(q^2_{\rm max})$
          & $F_{\rm HQ}(q^2_{\rm max})$
          & $a$
          & $b$
          & $F^{\P_Q\P_{Q'}}$
          & $F(0)$
          & $F(q^2_{\rm max})$
          & $F_{\rm HQ}(q^2_{\rm max})$
          & $a$
          & $b$
          \\
\hline     $f_1^{\Sigma'_{5b}\Sigma'_{5c}}$
          & 0.48
          & 0.97
          & 1
          & 1.62
          & 1.00
          & $g_1^{\Sigma'_{5b}\Sigma'_{5c}}$
          & 0.47
          & 0.92
          & 1
          & 1.53
          & 0.96
          \\
          $f_2^{\Sigma'_{5b}\Sigma'_{5c}}$
          & $-0.26$
          & $-0.58$
          & 0
          & 2.12
          & 1.91
          & $g_2^{\Sigma'_{5b}\Sigma'_{5c}}$
          & $-0.06$
          & $-0.13$
          & 0
          & 2.38
          & 2.56
          \\
\hline    $f_1^{\Theta_b\Theta_c}$
          & 0.45
          & 1.09
          & 7/6
          & 2.60
          & 3.23
          & $g_1^{\Theta_b\Theta_c}$
          & $-0.15$
          & $-0.31$
          & $-1/3$
          & $1.78$
          & $1.75$
          \\
          $f_2^{\Theta_b\Theta_c}$
          & 0.60
          & 1.55
          & 3/2
          & 2.59
          & 2.60
          & $g_2^{\Theta_b\Theta_c}$
          & $-0.05$
          & $-0.11$
          & 0
          & $3.48$
          & $7.18$
          \\
\hline
          $f_1^{\Sigma'_{5c} N^+_8}$
          & 0.80
          & 2.12
          & --
          & 2.10
          & 0.91
          & $g_1^{\Sigma'_{5c} N^+_8}$
          & 0.69
          & 1.43
          & --
          & 1.50
          & 0.72
          \\
          $f_2^{\Sigma'_{5c} N^+_8}$
          & $-0.91$
          & $-1.93$
          & --
          & 2.08
          & 2.77
          & $g_2^{\Sigma'_{5c} N^+_8}$
          & $-0.17$
          & $-0.36$
          & --
          & 3.28
          & 7.36
          \\
\hline    $f_1^{\Theta_c\Theta}$
          & 0.75
          & 1.37
          & --
          & 2.48
          & 5.51
          & $g_1^{\Theta_c\Theta}$
          & $-0.23$
          & $-0.48$
          & --
          & 1.83
          & 0.94
          \\
          $f_2^{\Theta_c\Theta}$
          & 1.15
          & 1.80
          & --
          & 2.30
          & 6.89
          & $g_2^{\Theta_c\Theta}$
          & $0.0050$
          & $0.0046$
          & --
          & $-1.34$
          & 2.09
\end{tabular}
\end{ruledtabular}
\end{table}

The $\P_Q\to\P_{Q'}$ transition form factors $f_{1,2}(q^2)$ and
$g_{1,2}(q^2)$ are given in Table~\ref{tab:fg} and  shown in
Fig.~\ref{fig:figi}. It can be easily seen that in most cases
these form factors at the zero recoil point ($q^2=q^2_{\rm max}$)
agree with the HQ expectation Eqs.~(\ref{eq:3to3zerorecoil}) and
(\ref{eq:6to6zerorecoil}). However, the form factor $|f_2(q^2_{\rm
max})|$ for the $\Sigma'_{5b}\to\Sigma'_{5c}$ transition is larger
than the HQ expectation and this may indicate the importance of
$1/m_c$ corrections in this case. Note that the aforementioned
heavy quark relations are not applicable to $\Theta_c\to\Theta$
and $\Sigma'_{5c}\to N^+_8$ transitions as the SU(3) quantum
numbers of the final states are different from that of the
corresponding initial states. The parameters $a$ and $b$ are of
order unity for most of the entries in Table \ref{tab:fg}. Note
that if $F(q^2)$ is parametrized without including the factor of
$(1-q^2/M^2)$ in the denominator of Eq.~(\ref{eq:FFpara1}), $a$
and $b$ will become larger.
It is worthy remarking that although there is not much information
on the parameter $\beta_{23}$ in the diquark system, the
dependence of the Gaussian wave function width in transition form
factors is mild as diquarks behave as a spectator. For example,
when $\beta_{23}$ is increased from 0.38~GeV to 0.55~GeV, all the
form factors at $q^2=0$ are reduced at most by 15\% except for
$g_2^{\Theta_b\Theta_c}$ and $g_2^{\Theta_c\Theta}$ that get
changed by $-40\%$ and $+50\%$, respectively,  In all cases, the
modification of form factors  is smaller at zero recoil than that
at maximum recoil. However, $g_2^{\Theta_b\Theta_c}(0)$ and
$g_2^{\Theta_c\Theta}(0)$ are of order $10^{-2}$ and $10^{-3}$,
respectively, and hence their contributions to decay rates are
small. Therefore, a variation of $\beta_{23}$ by $50\%$ will
modify the transition rates at most by $30\%$.

Under the factorization approximation, the decay amplitudes for
color-allowed $\P_Q\to\P_{Q'}\pi,\,P_{Q'}\rho$ decays in $\bar
b\to \bar c$ transitions are given by~\cite{Cheng97a}
 \be
 {\cal A}(\P_Q\to\P_Q'\pi)&=&i\bar u'(A+B\gamma_5) u,
 \non\\
  {\cal A}(\P_Q\to\P_Q'\rho)&=&\bar u'\varepsilon^{*\mu}(A_1\gamma_\mu\gamma_5+A_2 P'_\mu\gamma_5
 +B_1\gamma_\mu+B_2 P'_\mu) u,
 \en
where
 \be
 A&=&\frac{G_f}{\sqrt2} V^*_{cb} V_{ud}\, a_1 f_\pi (M'-M)
 f_1(m_\pi^2),\non\\
 B&=&\frac{G_f}{\sqrt2} V^*_{cb} V_{ud}\, a_1 f_\pi (M+M')
 g_1(m_\pi^2),\non\\
 A_1&=&-\frac{G_f}{\sqrt2} V^*_{cb} V_{ud}\, a_1 f_\rho
 m_\rho\left[g_1(m_\rho^2)+g_2(m_\rho^2)\frac{M-M'}{M+M'}\right], \\
 A_2&=&-2\frac{G_f}{\sqrt2} V^*_{cb} V_{ud}\, a_1 f_\rho
 m_\rho\frac{g_2(m_\rho^2)}{M+M'},\non\\
 B_1&=&-\frac{G_f}{\sqrt2} V^*_{cb} V_{ud}\, a_1 f_\rho
 m_\rho\left[f_1(m_\rho^2)-f_2(m_\rho^2)\right],\non\\
 B_2&=&-2\frac{G_f}{\sqrt2} V^*_{cb} V_{ud}\, a_1 f_\rho
 m_\rho\frac{f_2(m_\rho^2)}{M+M'}, \non
 \en
with $V_{ij}$ the CKM matrix element,
$f_{\pi\,(\rho)}=131~(216)$~MeV the pion (rho) decay constant and
$a_1(\sim 1)$ the effective color-allowed Wilson coefficient.
Likewise, the decay amplitudes for $\bar c\to \bar s$ transitions
have similar expressions with $V^*_{cb} V_{ud}$ replaced by
$V^*_{ud} V_{cs}$.

The decay rates read \cite{Cheng97a}
 \be
 \Gamma(\P_Q\to
 P_{Q'}\pi)&=&\frac{p_c}{8\pi}\left[\frac{(M+M')^2-m_\pi^2}
 {M^2}|A|^2+\frac{(M-M')^2-m_\pi^2}{M^2}|B|^2\right],
 \non\\
 \Gamma(\P_Q\to
 P_{Q'}\rho)&=&\frac{p_c}{8\pi}\frac{E'+M'}{M}\left[2(|S|^2+|P_2|^2)
 +\frac{E_\rho^2}{m_\rho^2}(|S+D|^2+|P_1|^2)\right],
 \en
with
 \be
 S&=&-A_1,\qquad P_1=-\frac{p_c}{E_\rho}\left(\frac{M+M'}{E'+M'} B_1+M
 B_2\right),\non\\
 P_2&=&\frac{p_c}{E'+M'} B_1,\qquad
 D=-\frac{p_c^2}{E_\rho(E'+M')}(A_1-M A_2),
 \en
where $p_c$ is the c.m. momentum. The decay rates for the weak
decays $\Theta_b^+\to\Theta_c^0\pi^+,\,\Theta_c^0\rho^+$,
$\Theta_c^0\to\Theta^+\pi^-,\,\Theta^+\rho^-$, $\Sigma'^+_{5b}\to
\Sigma'^0_{5c}\pi^+,\Sigma'^0_{5c}\rho^+$ and $\Sigma'^0_{5c}\to
N^+_8\pi^-,N^+_8\rho^-$ are summarized in Table~\ref{tab:rate}.
Assuming $\tau(\Theta_b^+)\sim \tau(\Lambda_b)\sim 1.2\times
10^{-12}$~s and $\tau(\Theta_c^0)\sim \tau(\Lambda_c)\sim 2\times
10^{-13}$~s for the weakly decaying $\Theta_b^+$ and $\Theta_c^0$,
we find $\B(\Theta_b^+\to\Theta_c^0\pi^+)\sim 1\times 10^{-3}$ and
$\B(\Theta_c^0\to\Theta^+\pi^-)\sim 4\%$, which are consistent
with the intuitive estimate made in \cite{Leibovich}.

\begin{table}[t!]
\caption{\label{tab:rate} The decay rates (in units of
$10^{10}\,s^{-1}$) of $\P_Q\to \P_{Q'}\pi,\,\P_{Q'}\rho$
for$a_1=1$. }
\begin{ruledtabular}
\begin{tabular}{ccccc}
          Mode
          & $\Sigma^{\prime+}_{5b}\to\Sigma^{\prime0}_{5c}\pi^+$
          & $\Theta^+_b\to\Theta^0_c\pi^+$
          & $\Sigma^{\prime0}_{5c}\to N^+_8\pi^-$
          & $\Theta^0_c\to\Theta^+\pi^-$
          \\
\hline    $\Gamma(10^{10}\,s^{-1})$
          & 0.23
          & 0.12
          & 42.62
          & 21.61
          \\
\hline \hline
          Mode
          & $\Sigma^{\prime+}_{5b}\to\Sigma^{\prime0}_{5c}\rho^+$
          & $\Theta^+_b\to\Theta^0_c\rho^+$
          & $\Sigma^{\prime0}_{5c}\to N^+_8\rho^-$
          & $\Theta^0_c\to\Theta^+\rho^-$
          \\
\hline    $\Gamma(10^{10}\,s^{-1})$
          & 0.34
          & 0.16
          & 82.41
          & 27.74
\end{tabular}
\end{ruledtabular}
\end{table}

Finally, it is worth commenting that $\Theta_c^0$ can be produced
in $B$ decays via the dominant modes $B^+\to
\Theta_c^0\bar\Delta^+$ and $B^0\to\Theta_c^0\bar p\pi^+$
\cite{Rosner}. Theoretically, it is difficult to estimate their
branching ratios. Nevertheless, the measured branching ratios by
Belle for charmful baryonic $B$ decays \cite{Belle}, $\B(\ov
B^0\to\Lambda_c^+\bar p)=(2.2^{+0.6}_{-0.5}\pm0.3\pm0.6)\times
10^{-5}$ and $\B(B^-\to\Lambda_c^+\bar
p\pi^-)=(1.87^{+0.43}_{-0.40}\pm0.28\pm0.49)\times 10^{-4}$,
provide some useful cue. Since a production of the pentaquark
needs one more pair of $q\bar q$ compared to the normal baryon, it
is plausible to expect that the branching ratios of $B^+\to
\Theta_c^0\bar\Delta^+$ and $B^0\to\Theta_c^0\bar p\pi^+$ are at
most of order $10^{-6}$ and $10^{-5}$, respectively. Hence, they
may be barely reachable at $B$ factories. As for the production of
the light pentaquark $\Theta^+$ in the decay $B^0\to\Theta^+\bar
p$, for example, it will be greatly suppressed for two reasons.
First, a typical two-body baryonic $B$ decay, e.g. $B^0\to p\bar
p$, occurs only at the level of $10^{-7}$ \cite{CKcharmless}.
Indeed, the upper limit of $B^0\to p\bar p$ has been recently
pushed to the level of $2.7\times 10^{-7}$ by BaBar \cite{BaBar}.
Second, $B^0\to\Theta^+\bar p$ is Cabibbo suppressed relative to
the $p\bar p$ mode. Therefore, it will be hopeless to detect a
light pentaquark production in two-body or three-body baryonic $B$
decays.

\section{conclusions}
Assuming the two diquark structure for the pentaquark as proposed
in the Jaffe-Wilczek model, we study the weak transitions of heavy
pentaquark states using the light-front approach. The main
conclusions are as follows.
 \begin{enumerate}
 \item
In the Jaffe-Wilczek model, there exist parity-even antisextet and
parity-odd triplet heavy pentaquark baryons and they are all truly
exotic. This differs than the Karliner-Lipkin model where the
triplet pentaquark states are parity-even. It is very likely that
the heavy pentaquarks in the $\3bar_f$ representation are lighter
than the $\6bar_f$ ones owing to the lack of orbital exciation in
the latter.
 \item
The theoretical estimate of charmed and bottom pentaquark masses
is rather controversial. It is not clear if the ground-state heavy
pentaquark lies above or below the strong-decay threshold. If the
narrow state observed by the H1 experiment can be identified as an
even-parity charmed pentaquark with a mass of 3099 MeV, then the
diquark in the Jaffe-Wilczek picture should not be treated as a
point-like object and it can have sizable hyperfine interactions
with the antiquark of the pentaquark baryon. Consequently, the
effective mass of the diquark $[ud]$ will be smaller in $\Theta^+$
than in $\Theta_c^0$ or $\Theta_b^+$. If the H1 state is a chiral
partner of the yet undiscovered ground-state charmed pentaquark
with opposite parity, the latter may be below the $DN$ threshold
and can only be discovered by studying its weak decays. In the
case, the point-like diquark picture could be valid. At any rate,
it is important to check and confirm the H1 state from other
experiments.
 \item
The antisextet-antisextet and triplet-triplet heavy pentaquark
weak transition form factors are calculated using the light-front
quark model. The momentum dependence of the physical form factors
is determined by first fitting the form factors obtained in the
spacelike region to a 3-parameter function in $q^2$ and then
analytically continuing them to the timelike region.
 \item
 In the heavy quark limit, it is found that heavy-to-heavy pentaquark
transitions can be expressed in terms of three Isgur-Wise
functions $\zeta$, $\xi_1$ and $\xi_2$: The first two are
normalized to unity at zero recoil as required by heavy quark
symmetry, while $\xi_2(1)$ is found to be equal to 1/2 at maximum
momentum transfer, in accordance with the prediction of the
large-$N_c$ approach or the quark model. Therefore, the
light-front model calculations are consistent with the requirement
of heavy quark symmetry.
 \item
Numerical results for form factors and Isgur-Wise functions are
presented. Decay rates of the weak decays $\Theta_b^+\to\Theta_c^0
\pi^+(\rho^+)$, $\Theta_c^0\to\Theta^+ \pi^-(\rho^-)$,
$\Sigma'^+_{5b}\to \Sigma'^0_{5c}\pi^+(\rho^+)$ and
$\Sigma'^0_{5c}\to N^+_8\pi^-(\rho^-)$ with $\Theta_Q$,
$\Sigma'_{5Q}$ and $N_8$ being the heavy anti-sextet, heavy
triplet and light octet pentaquarks, respectively, are obtained.
For weakly decaying $\Theta_b^+$ and $\Theta_c^0$, the branching
ratios of $\Theta_b^+\to\Theta_c^0\pi^+$,
$\Theta_c^0\to\Theta^+\pi^-$ are estimated to be of order
$10^{-3}$ and $10^{-2}$, respectively.
 \end{enumerate}

 \vskip 2.5cm \acknowledgments We are grateful to Ting-Wai Chiu
and Xiao-Gang He for valuable discussions. This research was
supported in part by the National Science Council of R.O.C. under
Grant Nos. NSC92-2112-M-001-016, NSC92-2811-M-001-054 and
NSC92-2112-M-017-001.


\begin{thebibliography}{99}
 \newcommand{\bi}{\bibitem}

 \bi{LEPS} LEPS Collaboration, T. Nakano {\it et al.,}
      Phys. Rev. Lett. {\bf 91}, 012002 (2003).

 \bi{Diana}
      DIANA Collaboration, V.V. Barmin {\it et al.,}
      Phys. Atom. Nucl. {\bf 66}, 1715-1718 (2003).

 \bi{CLAS1}
      CLAS Collaboration, S. Stepanyan {\it et al.,}
      Phys. Rev. Lett. {\bf 91}, 252001 (2003).

 \bi{Saphir}
      SAPHIR Collaboration, J. Barth {\it et al.}
      Phys. Lett. B {\bf 572}, 127-132 (2003).


 \bi{ITEP}
      A.E. Asratyan, A.G. Dolgolenko and M.A. Kubantsev,
      hep-ex/0309042.

 \bi{CLAS2}
      CLAS Collaboration, V. Kubarovsky {\it et al.,}
      Phys. Rev. Lett. {\bf 92}, 032001 (2004).

 \bi{Hermes}
      HERMES Collaboration, A. Airapetian {\it et al.,}
      \pl B {\bf 585}, 213 (2004).

 \bi{SVD}
      SVD Collaboration, A. Aleev {\it et al.,}
      hep-ex/0401024.

 \bi{COSY}
      COSY-TOF Collaboration, M. Abdel-Bary {\it et al.,}
      hep-ex/0403011.


 \bi{ZEUS} ZEUS Collaboration, S. Chekanov {\it et al.,}
   hep-ex/0404008.

 \bi{width} S. Nussinov, hep-ph/0307357; R.A. Arndt, I.I.
 Strakovsky, and R.L. Workman, \pr C~{\bf 68}, 042201 (2003);
 R.N. Cahn and G.H. Trilling, \pr D {\bf 69}, 011501 (2004).

 \bi{NA49}
      NA49 Collaboration, C. Alt {\it et al.,}
      Phys. Rev. Lett. {\bf 92}, 042003 (2004).

 \bi{BES} BES Collaboration, J.Z. Bai {\it et al.,}
   hep-ex/0402012.

 \bi{HeraB} HERA-B Collaboration, K.T. Kn\"opfle {\it et al.,}
 hep-ex/0403020.

\bibitem{JW}
R.~L.~Jaffe and F.~Wilczek,
Phys.\ Rev.\ Lett.\  {\bf 91}, 232003 (2003).

 \bi{DPP} D. Diakonov, V. Petrov, and M. Ployakov, Z. Phys. A {\bf
 359}, 305 (1997).

 \bi{KL} M. Karliner and H.J. Lipkin, \pl B {\bf 575}, 249 (2003).

 \bi{Csikor} F. Csikor, Z. Fodor, S.D. Katz, and T.G. Kovacs, JHEP
 {\bf 0311}, 070 (2003).

 \bi{Sasaki} S. Sasaki, hep-lat/0310014.

 \bi{Chiu} T.W. Chiu and T.H. Hsieh, hep-ph/0403020.

\bi{Cohen} T.D. Cohen, hep-ph/0402056.

 \bi{Lambda1405} W. Melnitchouk {\it et al.,} \pr D {\bf 67},
 114506 (2003); Y.N. Nemoto {\it et al., ibid.} {\bf 68}, 094505
 (2003).

 \bi{Nussinov} S. Nussinov, hep-ph/0403028.

 \bi{H1} H1 Collaboration, A. Aktas {\it et al.,} \pl B {\bf 588},
 17 (2004).

 \bi{Nowak} M.A. Nowak, M. Praszalowicz, M. Sadzikowski, and J.
 Wasiluk, hep-ph/0403184.

 \bi{Stewart} I.W. Stewart, M.E. Wessling, and M.B. Wise,
 hep-ph/0402076.

 \bi{Leibovich} A.~K.~Leibovich, Z.~Ligeti, I.~W.~Stewart and M.~B.~Wise,
 Phys.\ Lett.\ B {\bf 586}, 337 (2004). 

 \bi{He} X.G. He and X.Q. Li, hep-ph/0403191.

 \bi{Huang} P.Z. Huang, Y.R. Liu, W.Z. Deng, X.L. Chen, and S.L.
 Zhu, hep-ph/0401191.

 \bi{KLheavy} M. Karliner and H.J. Lipkin, hep-ph/0307343.

 \bi{Zhang} A. Zhang, Y.R. Liu, P.Z. Huang, W.Z. Deng, X.L. Chen,
 and S.L. Zhu, hep-ph/0403210.

 \bi{PDG} Particle Data Group,  K. Hagiwara {\it et al.,} \pr D {\bf
 66}, 010001 (2002).

 \bi{Cheung} K. Cheung, hep-ph/0308176.

 \bi{Callan} C.G. Callan and I.R. Klebanov, \np B {\bf 262}, 365
 (1985).

 \bi{Oh} Y. Oh, B.Y. Park, and D.P. Min, \pr D {\bf 50}, 3350
 (1994); \pl B {\bf 331}, 362 (1994).

 \bi{MaThetac} B. Wu and B.Q. Ma, hep-ph/0402244.

 \bibitem{Stancu}
 F.~Stancu,
 Phys.\ Rev.\ D {\bf 58}, 111501 (1998). 

 \bi{privateChiu}
 T.~W.~Chiu and T.~H.~Hsieh,
 hep-ph/0404007.

 \bi{Jenkins} E. Jenkins, \pr {\bf D54}, 4515 (1996); {\sl ibid.}
{\bf D55}, 10 (1997).

 \bi{Neubert94} For a review of heavy quark effective theory and the nonperturbative HQET
parameters, see M. Neubert, Phys. Rep. {\bf 245}, 259 (1994); Int.
J. Mod. Phys. A {\bf 11}, 4173 (1996).

 \bi{Cheng97a} H.Y. Cheng, \pr D {\bf 56}, 2783 (1997).

 \bi{Rosner} J.L. Rosner, hep-ph/0312269.

 \bi{Armstrong} S. Armstrong, B. Mellado, and S.L. Wu,
 hep-ph/0312344.

 \bi{Browder} T.~E.~Browder, I.~R.~Klebanov and D.~R.~Marlow,
 Phys.\ Lett.\ B {\bf 587}, 62 (2004). 

 \bi{E791} E791 Collaboration, E.M. Aitala {\it et al.,} \prl {\bf
 81}, 44 (1998); \pl B {\bf 448}, 303 (1999).

 \bibitem{CCH}
 H.~Y.~Cheng, C.~K.~Chua and C.~W.~Hwang,
 Phys.\ Rev.\ D {\bf 69}, 074025 (2004). 

 \bibitem{Cheng97} H. Y. Cheng, C. Y. Cheung, and C. W. Hwang, Phys.
        Rev. D {\bf 55}, 1559 (1997).

 \bi{Jaus90} W. Jaus, \pr D {\bf 41}, 3394 (1990).

 \bibitem{deAraujo:1999cr}
            W.~R.~de Araujo, M.~Beyer, T.~Frederico, and H.~J.~Weber,
            J.\ Phys.\ G {\bf 25}, 1589 (1999). 

  \bibitem{Brodsky:1997de}
  S.~J.~Brodsky, H.~C.~Pauli, and S.~S.~Pinsky,
  Phys.\ Rept.\  {\bf 301}, 299 (1998).

 \bibitem{Jaus91}
     W.~Jaus,
     Phys.\ Rev.\ D {\bf 44}, 2851 (1991).

  \bibitem{Schlumpf}
 F.~Schlumpf,
 Phys.\ Rev.\ D {\bf 47}, 4114 (1993); {\bf 49},
 6246(E) (1994)].

 \bibitem{Pauli}
 H.~C.~Pauli,
 Nucl.\ Phys.\ Proc.\ Suppl.\  {\bf 90}, 259 (2000)
 [arXiv:hep-ph/0103106].

  \bibitem{IW89} N. Isgur and M. B. Wise, Phys. Lett. B {\bf 232},
                113 (1989); {\bf 237}, 527 (1990).

 \bi{IW91}  N.~Isgur and M.~B.~Wise, \np B {\bf 348}, 276 (1991).

 \bi{Georgi} H. Georgi, \np B {\bf 348}, 293 (1991).

\bibitem{Yan92} T. M. Yan, H. Y. Cheng, C. Y. Cheung, G. L. Lin, Y. C.
        Lin, and H. L. Yu, Phys. Rev. D {\bf 46}, 1148 (1992);
        {\bf 55}, 5851(E) (1997);
         M. B. Wise, Phys.
        Rev. D {\bf 45}, R2188 (1992); G. Burdman and J. Donoghue,
        Phys. Lett. B {\bf 280}, 287 (1992).


 \bi{Chow} C.K. Chow, \pr D {\bf 51}, 1224 (1995); {\bf 54}, 873
 (1996).

 \bi{Chengbottom} H.Y. Cheng, \pr D {\bf 56}, 2799 (1997).

 \bibitem{CC}
  H.~Y.~Cheng and C.~K.~Chua,
 \pr D {\bf 69}, 094007 (2004).


 \bi{CCHZ} H.Y. Cheng, C.Y. Cheung, C.W. Hwang, and W.M. Zhang,
           \pr D {\bf 57}, 5598 (1998).

 \bibitem{Jaus96}
       W.~Jaus,
       Phys.\ Rev.\ D {\bf 53}, 1349 (1996); {\bf 54},
       5904(E) (1996).

  \bi{Belle} Belle Collaboration, N. Gabyshev {\it et al.,} \prl
{\bf 90}, 121802 (2003); \pr D {\bf 66}, 091102 (2002); K. Abe
{\it et al.,} \prl {\bf 89}, 151802 (2002).

 \bi{CKcharmless} H.Y. Cheng and K.C. Yang, \pr D {\bf 66}, 014020
(2002).

 \bi{BaBar} BaBar Collaboration, B. Aubert {\it et al.,}
 hep-ex/0403003.




\end{thebibliography}
\end{document}